\documentclass[12pt]{article}
\usepackage[top=1in, bottom=1in, left=1in, right=1in]{geometry}
\usepackage{amsmath, amsthm, amssymb}
\usepackage{graphicx}
\usepackage{algorithm} 
\usepackage{color}
\usepackage{algpseudocode}
\usepackage{epstopdf}
\usepackage{overpic}
\usepackage{cancel}
\usepackage{rotating}
\usepackage{url}
\usepackage{caption}
\usepackage{color}
\usepackage{rotating}
\usepackage{multirow}
\usepackage{wrapfig}
\usepackage{mathtools}
\usepackage{setspace}
\usepackage{mathrsfs}
\usepackage{cite}
\usepackage{hyperref}
\pdfoutput=1
\hypersetup{
  pdfinfo={
    Title={Data-driven Discovery of Cyber-Physical Systems},
    Author={Ye Yuan, et. al.},
  }
}
\usepackage{pdfpages}
\usepackage{subeqnarray}
\usepackage{cases}
\usepackage{color}
\usepackage[bottom,flushmargin,hang,multiple]{footmisc}

\usepackage{amsfonts}

\usepackage{amssymb}
\usepackage{amsmath,bm}
\usepackage{graphicx}
\usepackage{subfigure}
\usepackage{array}
\usepackage{mathrsfs}
\usepackage{mathptmx}
\usepackage{color}
\usepackage{xmpmulti}
\usepackage{colortbl,dcolumn}
\usepackage{tcolorbox}
\usepackage{arydshln}
\usepackage{longtable}
\usepackage{makecell}
\usepackage{amsmath,amssymb}

\newcommand{\noise}{\boldsymbol{\xi}}

\usepackage{fancyhdr}
\usepackage{textcomp,xcolor}
\definecolor{MatlabCellColour}{RGB}{250,250,250}
\definecolor{MatPurp}{rgb}{.625,.1406,.9375}
\usepackage{listings}
 \usepackage[section]{placeins}

\lstdefinestyle{customc}{
  belowcaptionskip=.25\baselineskip,
  breaklines=true,
  frame=L,
  xleftmargin=\parindent,
  language=Matlab,
  showstringspaces=false,
  basicstyle=\small\ttfamily,
  keywordstyle=\bfseries\color{white!30!black},
  identifierstyle=\color{blue},
  commentstyle=\itshape\color{green!60!black},
  stringstyle=\color{MatPurp},
  backgroundcolor=\color{MatlabCellColour}
 }

\definecolor{blue}{rgb}{0,0,1}
\definecolor{darkgreen}{rgb}{0,0.5,0}
\definecolor{red}{rgb}{1,0,0}
\definecolor{teal}{rgb}{0,0.5,0.7}

\newenvironment{sciabstract}{%
\begin{quote} \bf}
{\end{quote}}

     \makeatother
\allowdisplaybreaks

\begin{document}

\title{\bf Data-driven Discovery of Cyber-Physical Systems}
\author
{Ye Yuan$^{1,2}$, Xiuchuan Tang$^{1,3}$, Wei Pan$^4$, Xiuting Li$^1$, Wei Zhou$^1$,\\ Hai-Tao Zhang$^{1,2}$, 
Han Ding$^{2,3,\ast}$, Jorge Goncalves$^{5,6,\ast}$\\
\normalsize{$^{1}$School of Automation, Huazhong University of Science and Technology}\\
\normalsize{$^{2}$State Key Lab of Digital Manufacturing Equipment and Technology}\\
\normalsize{$^{3}$School of Mechanical Science and Engineering,} \\\normalsize{Huazhong University of Science and Technology}\\
\normalsize{$^{4}$Department of Cognitive Robotics, Delft University of Technology}\\
\normalsize{$^{5}$ Department of Engineering, University of Cambridge
}\\
\normalsize{$^{6}$Luxembourg Centre for Systems Biomedicine, University of Luxembourg}\\
\normalsize{$^\ast$E-mail: dinghan@hust.edu.cn, jmg77@cam.ac.uk.}
}
\date{}
\maketitle
\begin{sciabstract}
Cyber-physical systems (CPSs) embed software into the physical world. They appear in a wide range of applications such as smart grids, robotics, intelligent manufacture and medical monitoring. CPSs have proved resistant to modeling due to their intrinsic complexity arising from the combination of physical components and cyber components and the interaction between them. This study proposes a general framework for reverse engineering CPSs directly from data. The method involves the identification of physical systems as well as the inference of transition logic. It has been applied successfully to a number of real-world examples ranging from mechanical and electrical systems to medical applications. The novel framework seeks to enable researchers to make predictions concerning the trajectory of CPSs based on the discovered model.  Such information has been proven essential for the assessment of the performance of CPS, the design of failure-proof CPS and the creation of design guidelines for new CPSs. 
\end{sciabstract}

\section*{Introduction}
\noindent Since the invention of computers, software has quickly become ubiquitous in our daily lives. Software controls domestic machines, such as washing and cooking appliances, aerial vehicles such as quadrotors, the scheduling of  power generation and the monitoring of human body vital signals. These technologies embed cyber components throughout our physical world. In fact, almost all modern engineering systems involve the integration of cyber and physical processes. \\ \\
The integration of cyber and physical components provides new opportunities and challenges. On one hand, this integration produces new functionality in traditional physical systems, such as brakes and engines in vehicles, intelligent control systems for biochemical processes and wearable devices {\it\cite{cps,hybridsysintro,Aihara2010royal}}. On the other hand, the integration of cyber components adds new layers of complexity, potentially seriously complicating their design and guaranteeing their performance. 
CPSs, such as modern power grids or autonomous cars, require guarantees on performance to be economically and safely integrated into society. In power grids, the failure of transformer taps, capacitors and switching operations alter the dynamics of the grid. Changes in dynamics of the power grid can be extremely costly. We have, after all, already witnessed a massive power outage in Southern California on September 2011 due to a cascading failure from a single line tripping (which was not detected by operators using their model), costing billions of USD. In autonomous driving, when operating in multiple complex scenarios -- from driving on multi-lane highway to turning at intersections while obeying rules -- high-level software makes decisions while low-level computer control systems realize the command using a combination of GPS/IMU, camera, radar and LIDAR data {\it\cite{ad}}. In such complex scenarios, guaranteeing CPS's performance poses a fundamental challenge.\\  \\
For performance guarantees, we require reliable models that capture essential dynamics. 
The central question this study seeks to answer, therefore, is how to reliably and efficiently automate mechanistic modeling of CPSs from data {\it\cite{kutz,Wang2016pr}}. An appropriate mathematical model of CPS should recognize the hybridity of CPS, which comprise of discrete and continuous components due to the integration of software and physical systems, respectively. Hybrid dynamical systems (detailed in the Materials and Methods section below and Supplementary Materials Section~S1.2) use finite-state machines to model the cyber components and dynamical systems for the physical counterparts. Hybrid dynamical models can produce accurate predictions and enable assessments of the CPS's performance {\it\cite{hybridlecture}}. This paper presents a new method, namely identification of hybrid dynamical systems (IHYDE), for automating the mechanistic modeling of hybrid dynamical systems from observed data and without any prior knowledge. IHYDE has low computational complexity and is robust to noise, enabling its application to real-world CPS problems. \\ \\
There are various methods for identifying nonhybrid dynamical systems. Schmidt and Lipson {\it\cite{Schmidt2009science}} proposed a data-driven approach to determine the underlying structure and parameters of time-invariant nonlinear dynamical systems. Schmidt and Lipson's method uses symbolic regression to identify the system, balancing model complexity and accuracy. However, symbolic regression has its limitations: it is computationally expensive, does not scale to large systems, and is prone to overfitting. Although recent research {\it\cite{Pan2012cdc, Wang2011prl, Chang2012cdc, Brunton2016pnas}} managed to reduce the expensive computational burden using compressive sensing and sparse learning, these methods cannot be applied to hybrid dynamical systems because of the complexity in hybrid models; basically, these algorithms cannot account for an unknown number of unknown subsystems that interact via unknown transition logic. \\ \\
There has been a number of interesting results in hybrid dynamical system identification in the past two decades  {\it\cite{hybrididljung,hybridid1,hybridid,hybridid2,bako,hybridid8,hybridid3,hybridid4,hybridid5,hybridid6,hybridid7,Oishi,bayen}}. Researchers have been developing different methods across several fields such as algebraic-geometric {\it\cite{hybridid2}}, mixed integer programming {\it\cite{hybridid3}}, bounded-error  {\it\cite{hybridid4}}, Bayesian learning {\it\cite{hybridid5}}, clustering-based strategies {\it\cite{hybridid6}}, and multi-modal symbolic regression {\it\cite{hybridid7}}. Reference {\it\cite{hybridid}} gives a comprehensive literature review, which summarizes all major progresses at the time. Later, pioneering works {\it\cite{bako,hybridid8}} use ideas from compressive sensing {\it\cite{Candes2006picm}} to identify the minimum number of submodels by recovering a sparse vector-valued sequence. Despite the clear merits of all these pioneering contributions, yet most research on hybrid system identification has been dealing with the most basic hybrid dynamical model-- the piecewise affine model with linear transition rules {\it\cite{hybridid1}}. These methods require some type of prior knowledge of the hybrid system, such as number of subsystems, parametrization of subsystems dynamics or transition logic. In contrast, IHYDE removes all these assumptions, and with no prior knowledge of the system (except perhaps the general field of the system), provides the number of subsystems, their dynamics, and the transition logic. IHYDE deals with this problem in two parts: first, the algorithm discovers how many subsystems are interacting -- and identifies a model for each one; second, the algorithm infers the transition logic between each pair of subsystems. Later in this work, we will propose IHYDE and detail the two-step method for discovering hybrid dynamical systems from data directly. Next, we present the results of using IHYDE on a number of examples, ranging from power engineering and autonomous driving to medical applications, to demonstrate the algorithm's application to various types of datasets.

\section*{Results}
This section is divided in two major parts. The first presents the proposed inference-based IHYDE algorithm using a simple example-- a thermostat, while the second illustrates its applicability to a wide range of systems, from real physical systems to challenging in silico systems, and from linear to nonlinear dynamics and transition rules. Details of both the algorithm and how data was acquired or generated can be found in Materials and Methods or Supplementary Materials.

\subsection*{The inference-based IHYDE algorithm applied to a thermostat}

\begin{figure}[!]
\centering
	\includegraphics[scale=.5]{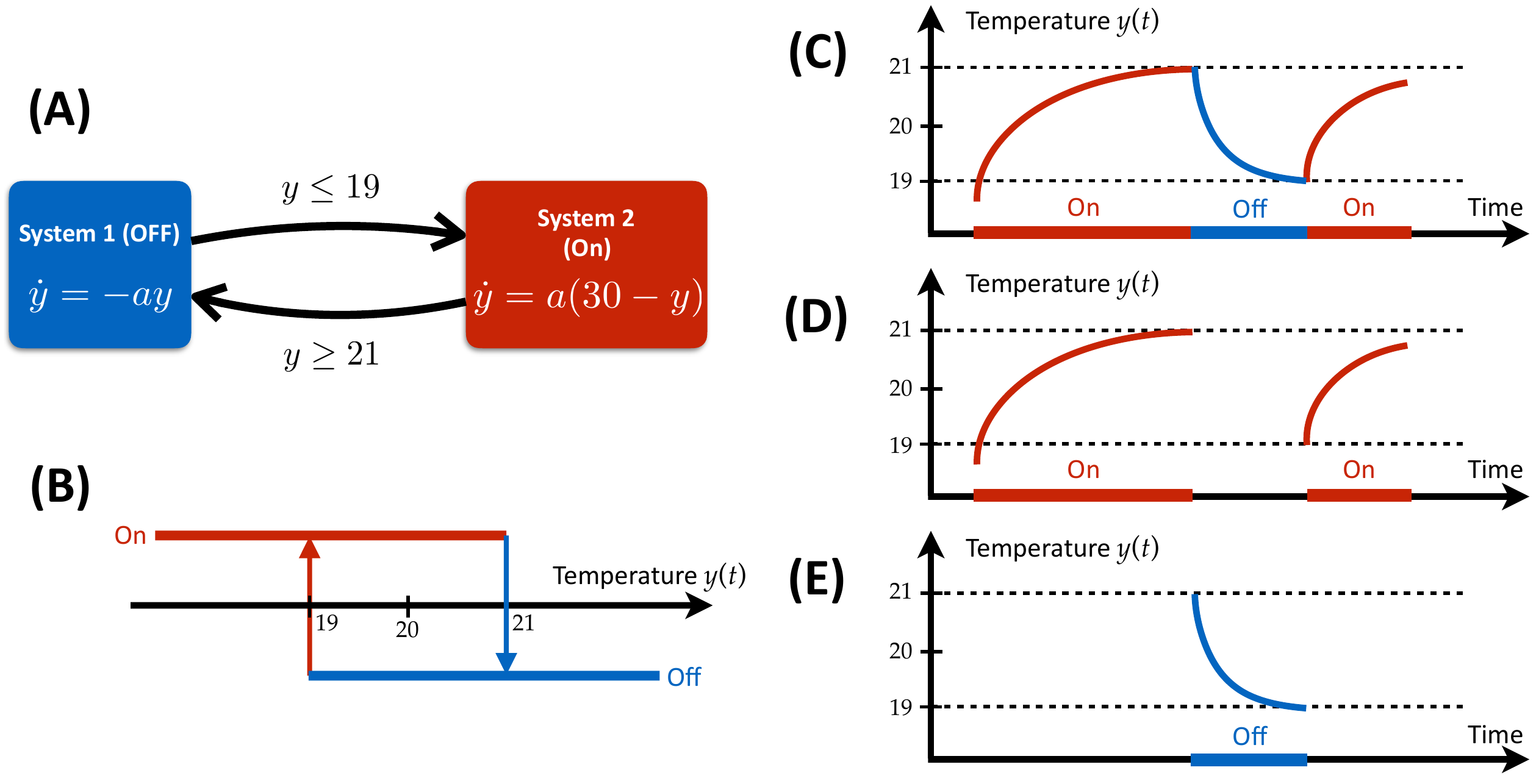}
	\caption{Thermostat example. {\bf (A)} The physical dynamic equations plus the transition rules of the hybrid system. {\bf (B)} The transition rules of the relay hysteresis. {\bf (C)} A simulation of the temperature of the thermostat system. Red (blue) is associated with the heater on (off). {\bf (D)} {\bf (E)} Values of the temperature corresponding to the heater on (off).}
	\label{fig:ex_ts}
\end{figure}

This section explains the key concepts of IHYDE using one of the simplest and ubiquitous hybrid dynamical systems: a room temperature control system consisting of a heater and a thermostat. The objective of the thermostat is to keep the room temperature $y(t)$ near a user specified temperature. At any given time, the thermostat can turn the heater on or off. When the heater is off, the temperature dissipates to the exterior at a rate of $-ay(t)$ degrees Celsius per hour, where $a>0$ is related to the insulation of the room. When the heater is on, it provides a temperature increase rate of $30a$ degrees Celsius per hour (Fig.~\ref{fig:ex_ts}A).\\ \\
Assume a desired temperature is set to $20$ degrees Celsius. Thermostats are equipped with hysteresis to avoid chattering, i.e., fast switching between on and off. A possible transition rule is to turn the heater on when the temperature falls bellow $19$ degrees, and switching it off when it reaches $21$ degrees (Fig.~\ref{fig:ex_ts}B). The goal of IHYDE is to infer both subsystems plus the transition logic from only the observed time-series data of the temperature (Fig.~\ref{fig:ex_ts}C).

\subsubsection*{Inferring subsystems}
The first step of the algorithm is to iteratively discover which subsystem of the thermostat generated which time-series data. 
Initially, the algorithm searches for the subsystem that captures the most data, since this subsystem would explain the largest amount of data. In this case, the algorithm would firstly find subsystem $2$ (heater on) since more than half of the data corresponds to that subsystem (Fig.~\ref{fig:ex_ts}C). The time-series portion of the data (Fig.~\ref{fig:ex_ts}D) is then used to find the dynamics of subsystem 2. The algorithm is then repeated on the remaining data (Fig.~\ref{fig:ex_ts}E). In this case, there is only one subsystem left (heater off). Hence, the algorithm classified all the data to a subsystem and identified the corresponding dynamics.


\subsubsection*{Inferring transition logic}

The second and final step is to identify the transition logics between the two subsystems, i.e. what triggered the transitions from on to off and from off to on. Starting with subsystem 2 (heater on) and its associated data in Fig.~\ref{fig:ex_ts}D, the algorithm first learns that no switch occurs when the temperature changes from just below $19$ to near $21$. Since the switch happens when the temperature reaches $21$ degrees, the algorithm concludes that the switch from on to off happens when $y(t)=21$ degrees. In practice, however, the software detects the switches when $y(t) \geq 21$. Similarly, from Fig.~\ref{fig:ex_ts}E, the algorithm learns that the switch from on to off happens when $y(t) \leq 19$.  In summary, IHYDE automatically learns the dynamics of all subsystems and the transition rules from one subsystem to another. While this is a simple system, as we will show next, this is true even in the presence of a large number of subsystems, potentially with nonlinear dynamics and transition rules.

\subsection*{Universal application}
Next, we illustrate how IHYDE can be applied a wide range of applications, from power engineering to robotics to medicine, showing the flexibility, applicability and power of IHYDE to model complex systems. Here, we consider the following systems. 1) Autonomous vehicles and robots: design and validation of an autonomous vehicle. 2) Large scale electronics: Chua'a circuit. 3) Monitoring of industrial processes: monitoring a wind turbine. 4) Power systems: transmission lines and smart grids. 5) Medical applications: heart atrial AP monitoring. To test IHYDE's performance, these systems will include both real experiments and synthetic datasets. Details can be found below and in Section~S3 of Supplementary Materials. \\ \\
Table~\ref{table:main} contains a summary of the most important systems analyzed in the paper. The first three examples are based on real data, while the other three are based on simulated data. The first column illustrates the systems, while the second column shows the different subsystems plus the transition rules. Each subsystem is associated with a particular color. The third column shows the original time-series data (dots) in the color associated with the subsystem that generated it, the fitted data from the identified models (lines connecting the dots), and the location of the transitions (changes in colors). Note that, at this resolution, the original data and the data obtained from the fitted models are indistinguishable. The last column presents the relative error ratio {\it\cite{Ljung1999book}} between the true data and the data simulated by the fitted model. A small error ratio indicates a good agreement between the true and modeled systems, and serves as a measure of the performance of IHYDE. Data is either collected (real systems) or simulated (synthetic systems) and captures all key transitions. As seem in column 3 and column 4 of Table~\ref{table:main}, IHYDE successfully modeled the original dynamics that generated the data in all examples with extremely high precision (nearly zero identification errors). First, it was able to classify each time point according to the respective subsystem that generated it. Second, it identified the dynamics of each subsystem with a very small error (less than 0.3\% on all simulated examples). Finally, it correctly identified the transition rules between subsystems.

\begin{table}[h]
\centering
	\includegraphics[scale=0.9]{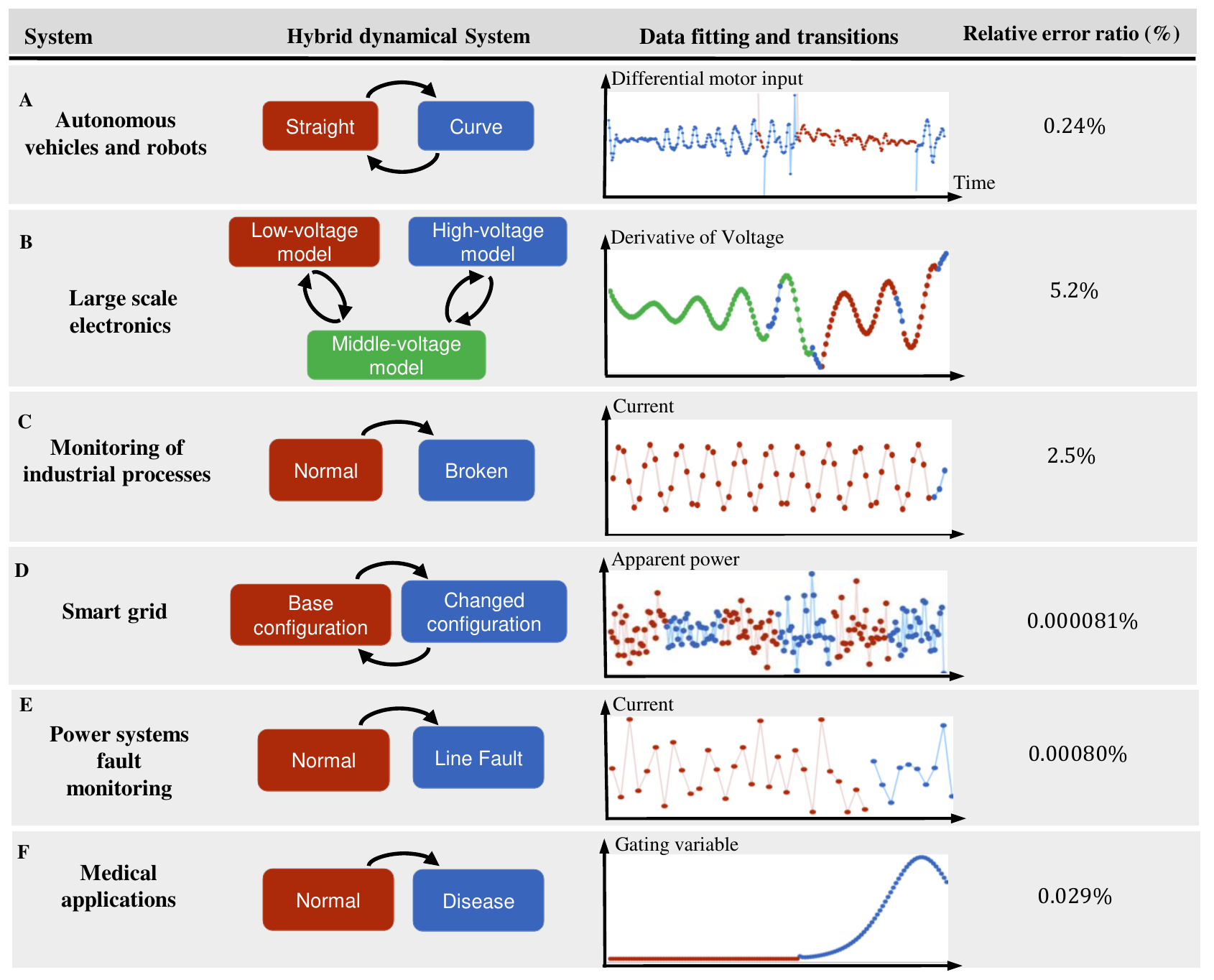}
	\caption{Summary of IHYDE algorithm applied to numerous examples.}
	\label{table:main}
\end{table} 

\subsubsection*{Autonomous vehicles and robots: design and validation}
To demonstrate IHYDE's usefulness in designing and validating complex systems, we tested the algorithm on an  autonomous vehicle, custom built in the lab (Table~\ref{table:main}A). Typically, the design process of complex systems consists of an arduous, time-consuming, and trial-and-error based approach: start from an initial design, evaluate it's performance and revise it until the performance is satisfied. A primary issue with this iterative approach is that when a design fails to meet desired specifications, many times engineers have little to no insight on how to improve the next iteration. Often, an engineer cannot discern whether the issue is due to poor mechanical design, issues with the software, or factors that were not considered. And this is also true with other general complex systems that involve interactions between physical/mechanical parts and software.\\ \\
%
%
The autonomous electrical car consists of a body, a MK60t board, a servo motor, a driving motor, and a camera. The design goal of the autonomous car was to successfully run through a winding track as quickly as possible. Using an embedded camera, the software captures information of the upcoming road layout to ascertain whether a straightway or a curve is coming up. Based on this information, the motor chooses an appropriate power to match the desired speed control strategy. For the purpose of illustration, we considered a simple controller that provides higher velocities on straightways and lower velocities on curves. In addition, simple feedback controllers help the car follow the chosen speed and stay on the track. The speed control strategy is based on incremental proportional and integral (PI) control that keeps the car at the correct speed, while the switching rule decides on the correct speed depending on whether a straight or curve is coming up. \\ \\
For the first design, we deliberately swapped the straightway and curve speeds to mimic a software bug. As a consequence, the car travelled rather slow in the straights and left the tracks in the curves (Supplementary Materials: movie S1). While in this case it was rather easy for engineers to spot the software bug, debugging, in general, can be extremely difficult, sometimes only possible by trial and error, and, as a consequence, very time consuming. One would like to check whether these types of bugs could be pinpointed by IHYDE. Indeed, from the data generated by the faulty system, IHYDE showed that the models had incorrect speed controllers. Hence, from data alone, IHYDE successfully reverse engineered the control strategy of the CPS.

\subsubsection*{Large scale electronics (Chua's circuit)}  
Debugging and verifying large scale electronics can be a daunting experience. Modeling could help identify whether a device has been built according to the desired specifications by identifying faulty connections or incorrect implementations. Simple electrical circuits, such as RLC circuits, are linear and easy to model. However, most electronic circuits introduce both nonlinear dynamics and switches (e.g., diodes and transistors), which can lead to extremely complex behaviors. Thus, modeling such systems can be very challenging.\\ \\
To illustrate IHYDE in this scenario, we built an electronic circuit that exhibits complex behaviors. We chose a well known system, called the Chua's circuit {\it\cite{chua}}, that exhibits chaotic trajectories (Table~\ref{table:main}B). Chaotic systems constitute a class of systems that depend highly on initial conditions, and makes simulation and modeling very challenging. Our circuit consists of an inductor, two capacitors, a passive resistor and an active nonlinear resistor, which fits the condition for chaos with the least components. The most important active nonlinear resistor is a conceptual component that can be built with operational amplifiers and linear resistors. The resulting nonlinear resistor is piecewise linear, making the Chua's circuit a hybrid dynamical system with a total of three subsystems.\\ \\
After collecting real data measured from the circuit, IHYDE successfully captures the dynamics of system and the transition rules between identified subsystems. In particular, the nonlinear dynamics are consistent with the true parameters of the circuit elements. As with all examples, modeling of the Chua's circuit was achieved using only the data, and no other assumptions on dynamics or switching behaviors. 

\subsubsection*{Monitoring of industrial processes (wind turbines)}
Next, we consider the problem of real-time monitoring industrial processes. Modeling large scale industrial processes is challenging due to the large number of parts involved, nonlinear dynamics and switching behaviors. Switches, in particular, are caused by breaking down of parts (due to wear and tear) and turning processes on and off, which introduce discontinuities in the dynamics.  We propose IHYDE as a tool to detect these switches as quickly as possible to prevent lengthy and expensive downtimes in industrial processes.\\ \\
To put IHYDE to the test, we used real data from a wind turbine platform built in {\it\cite{HeQun2017}}. We measured the current generated by the wind turbine under different operating conditions (Table~\ref{table:main}C). The system included a 380V power supply, a variable load, a power generator, a motor, a fan, two couplings and a gearbox that transmits the energy generated by the wind wheel to the power generator {\it\cite{HeQun2017}}. We performed experiments under normal and faulty conditions (a broken tooth of gearbox (Fig.~S14)) and measured the current of the wind turbine with sampling frequency $1000$Hz. In both experiments, the generator speed was $200$ revolutions per minute and the load was $1.5$ $\rm KNm$. \\ \\
IHYDE was tested under two different scenarios: offline modeling, used, for example, at the design stage; and online modeling, for real-time monitoring. In offline modeling, all the data are available for modeling, while in real-time monitoring only past data are available, and the system is continuously modeled as new data is gathered. In offline modeling, IHYDE identifies two linear subsystems, corresponding to the system in the two different conditions. In addition, it correctly detects the fault right after it happens and infers the transition logic. In online modeling, a model predicts the next time-series data point, and compares it with the real one, when this becomes available. If the difference is high, IHYDE detects a transition, builds a new model, and compares it with the old model to pinpoint the location of the fault. In this example, we focus on the online modeling: the fault is detected within only $3$ data points, or within $1$ sec, following its occurrence. This application demonstrates the capabilities of IHYDE in online monitoring of industrial processes. 

\subsubsection*{Power systems (smart grids and transmission lines)}
Smart grids have been gaining considerable attention in the last decades and are transforming how power systems are developed, implemented and operated.  They considerably improve efficiency, performance and makes renewable power feasible. In addition, it overhauls aging equipment and facilitates real-time troubleshooting, which decreases brownouts, blackouts, and surges.  As with all critical infrastructures, smart grids require strict safety and reliability constraints. Thus, it is of great importance to design monitoring schemes to diagnose anomalies caused by unpredicted or sudden faults {\it\cite{Pan2015Automatica}}. Here, we consider two examples of power systems: real-time modeling to control smart grids and pinpointing the location of a transmission line failure. \\ \\
We start by illustrating how IHYDE can model and control smart grids in real-time. Accurate model information is not only necessary for daily operation and scheduling, but also critical for other advanced techniques such as state estimation and optimal power flow computation. However, such information is not always available in distribution systems due to frequent model changes. These changes include: high uncertainty in distributed energy resources, such as components being added and removed from the network; unexpected events, such as line faults and unreported line maintenance; and trigger of automatic control and protection measures.
 We apply IHYDE to identify network models and infer transition logics, capturing model changes from advanced metering infrastructure data and in real-time. 
The 33-bus benchmark distribution system {\it\cite{baran1989network}}, shown in Figure~S8, generates the data. It is a hypothetical 12.66 KV system with a substation, 4 feeders, 32 buses, and 5 tie switches {\it\cite{baran1989network}}. The system is not well-compensated and lossy, and is widely used to study network reconfiguration problems.
Assume the loads on some remote nodes of a feeder suddenly increase, causing voltage sag. Subsequently, an operator takes switch action for load balancing and voltage regulation. Figure~S9 depicts the switching topologies and the real transition logic. 
Suppose we can measure all active and reactive power consumption, and voltage phasors of the nodes. Hence, the system is changing between two configurations corresponding to topologies when some switches turn on and off. For each node and subsystem, IHYDE successfully identifies the responding column of the admittance matrices with nearly zero identification errors. The identified admittance matrices at the switching time instants are very different from that of the previous moments, indicating a model switching (corresponding to changes in colors on the data in Table~\ref{table:main}D). Indeed, the identified logic is consistent with the real logic and demonstrates that IHYDE can reveal voltage drops at specific nodes in real-time and suggest switch action to avoid sharp voltage drops.\\ \\
To simulate a transmission line failure, assume a transmission line fails between two buses in the network. We will use a standard benchmark IEEE 14-bus power network (please see {\it\cite{power_link}} and Fig.~S7). This system consists of  generators, transmission lines, transformers, loads and capacitor banks. Looking directly at the generated data (Table~\ref{table:main}E), it is not clear when the fault occurred, and much less what happened at the time of failure and where it was located. This is because the power system compensated the failure by rerouting power across other lines. IHYDE, however, can immediately detect the occurrence of this event and determine its location. This is done by estimating the new admittance matrix using only 10 measurements following the failure (corresponding to 166.7 milliseconds, according to the IEEE synchrophasor measurements standard C37.118, 2011). Basically, it successfully discovers both subsystems (normal and failure) from data and calculates the difference of the discovered subsystems (leading to the location of the fault).
Given the frequency at which Phasor Measurement Units (PMUs) sample voltage and current, IHYDE is able to locate the fault in a few hundred milliseconds after the event occurs, enabling the operators to detect the event, identify its location, and take remedial actions in real-time.

\subsubsection*{Medical applications (heart atrial AP monitoring)}
The development of medical devices is another active research area. Especially, with the widespread use of wearables and smart devices, there is an exponential growth of data collection. These data requires personalized modeling algorithms to extract critical information for diagnosis and treatments. Within this context, we apply IHYDE to model data gathered from a human atrial action potential (AP) system {\it\cite{courtemanche1998ionic}}. The human atrial AP and ionic currents that underlie its morphology are of great importance to our understanding and prediction of the electrical properties of atrial tissues under normal and pathological conditions.\\ \\
The model captures the spiking of the atrial AP. In particular, two gating variables capture the fast and slow inactivation with switching dynamics. Following a spike, these two variables raise, preventing a new spike. Eventually, as the AP returns to low values, the inactivation dynamics switch back, and in time allow a new spike to take place. The goal is to test whether IHYDE can detect these transitions, together with the rules that led to the switch. In this study, IHYDE indeed identifies the two subsystems, together with their dynamics, and pinpoints the changing logic correctly (Table~\ref{table:main}F).  Hence, IHYDE provides a reliable model to study the system and to build devices to detect abnormal AP.

\section*{Discussion}
This work presents a new algorithm for identifying CPS from data. The algorithm does not require any prior knowledge and assumptions, except perhaps the general area of the system (e.g. a power grid or a biological system). IHYDE successfully identifies complex mechanistic models directly from data, including the subsystem dynamics and their associated  transition logics. The proposed method differs from classical machine learning tools, such as deep neural network models {\it\cite{dl}}, which typically do not provide insight on the underlying mechanisms of the systems. 
While IHYDE is inspired by prior work in symbolic regression  {\it\cite{Ly2012jmlr}}, it has much lower computational complexity due to the use of sparse identification and artificial intelligence. As a result, it can solve large-scale CPSs, facilitating its application to complex real-world problems.\\ \\
After IHYDE models a CPS, the resulting model can help verify the design specifications and predict future trajectories. If the CPS model reveals design flaws or fails to meet desired requirements, it can guide the redesign to achieve the required performance. Applications include robotics and automated vehicles (such as cars and unmanned air/spacecraft), where data-driven models promise to reduce the reliance on trial and error. Furthermore, IHYDE can monitor, detect, and pinpoint real-time faults of CPSs (for example, power systems), thereby helping avoid catastrophic failures.
%
%
%
%
As seen in the results section, IHYDE can be applied to a wide range of applications. Supplementary Materials includes additional examples on canonical hybrid dynamical systems {\it\cite{Ly2012jmlr}}. As before, IHYDE successfully identifies both the subsystems and the transition rules with virtually zero error (see Example 1-4 in Supplementary Materials).\\ \\
IHYDE unifies previous results as it can discover not only hybrid dynamical systems, but also nonhybrid dynamical systems (i.e., time-invariant linear and nonlinear systems {\it\cite{Brunton2016pnas, Pan2012cdc}}) as special cases. This was confirmed in Supplementary Materials Section~S2, where IHYDE successfully identified the original canonical dynamical systems from the data in {\it\cite{Brunton2016pnas}} (Table~S31). Hence, IHYDE provides a unified approach to the discovery of hybrid and non-hybrid dynamical systems.
\\ \\
While the approach has a number of advantages, there are still some open questions. First, it requires a new theory to understand when particular datasets are informative enough to uniquely identify a single (the true) hybrid dynamical system. Identifiability is a central topic in system identification and provides guarantees that there does not exist multiple systems that can produce the same data. This is illustrated in Supplementary Materials Section~S4, where we construct several hybrid dynamical systems that yield the exact same data, and hence cannot be differentiated from data alone. A second issue is on the choice of a suitable tuning parameter of IHYDE that balances model complexity and fitness in the identification process. This often requires fine-tuning and cross validation; the result will vary considerably according to decision made. 

\subsection*{Materials and Methods}
\subsubsection*{Hybrid dynamical systems}
A formal definition of hybrid dynamical systems can be found in {\it\cite{Ly2012jmlr,hybridanalysis}} and in Supplementary Materials. Here, we summarize the key aspects. Physical systems are characterized by inputs $u(t) \in\mathbb{R}^m$ and outputs $y(t)\in\mathbb{R}^n$. Based on these variables, at any given time  a particular mode $m(t)$ is chosen from a possible total of $K$ modes, i.e., $m(t) \in \{1,2,...,K\}$. Each mode corresponds to particular sets of physical parameters. \\ \\
The physical system evolves according to sets of differential equations: 
$$\frac{dy(t)}{dt}=\mathbf{F}_k\left(y(t),u(t)\right), \ \ \ \ \  k=1,2,\ldots,K,$$
where each $\mathbf{F}_k(y(t),u(t)))$ is related to the dynamics of subsystem $k$. Assume $y(t)$ and $u(t)$ are sampled at a rate $h>0$, i.e.  sampled at times $0, h, 2h, 3h...$. For fast enough sampling (for small sampling period $h$), one of the simplest method to approximate derivatives is to consider
$$\frac{dy(t)}{dt} \approx \frac{y(t+h)-y(t)}{h},$$
which yields the discrete-time system
$$y(t+h) = y(t) + h\ {\bf F}_k(y(t),u(t)) \triangleq {\bf f}_k(y(t),u(t)), \ \ \ \ \  k=1,2,\ldots,K.$$
At any given time, the decision of the transition logic to switch to another subsystem is governed transition rules of the form
$$m(t+h) = \mathcal{T}(m(t), y(t), u(t)).$$
Hence, the current input-output variables $y(t), u(t)$ plus the current subsystem $m(t)$ determine, via a function $\mathcal{T}$, the next subsystem. Without loss of generality, we can rescale the time variable $t$ so that $h=1$. Thus, we can construct the following mathematical model for hybrid dynamical systems
\begin{equation*}
\begin{aligned}
m(t+1) & = \mathcal{T}(m(t), y(t), u(t))\\
y(t+1) &= {\bf f}(m(t),y(t), u(t)) = \left\{
\begin{aligned}
{\bf f}_1(y(t), u(t))&,& \text{if}~m(t) = 1\\ 
\vdots ~& , & \vdots\\
{\bf f}_K(y(t), u(t))&,& \text{if}~m(t) = K
\end{aligned}
\right.
\end{aligned}
\end{equation*}

%

%
\subsubsection*{Subsystem identification}
Let $Y$ and $U$ denote column vectors containing all the samples of $y(t)$ and $u(t)$, respectively, for  $t=1,2,\ldots, M+1$, where $M+1$ is the total number of samples. The first step to identify the subsystems is to construct a library $\Phi(Y,U)$ of nonlinear functions from the input-output data. The exact choice of nonlinear functions in this library depends on the actual application.  For example, for simple mechanical systems, $\Phi$ would consist of constant, linear and trigonometric terms; in biological networks, $\Phi$ would consist of polynomial (mass action kinetics) and sigmoidal (enzyme kinetics) terms. Let
\begin{eqnarray*}
Y = \begin{bmatrix} 
~~\vline&\vline &  \vline & \vline~~  \\
~~y(1)& y(2)   & \ldots & y(M) ~~\\
~~\vline &\vline & \vline &\vline~~ 
\end{bmatrix} ^T, ~~U = \begin{bmatrix} 
~~\vline&\vline &  \vline & \vline~~  \\
~~u(1)& u(2)   & \ldots & u(M) ~~\\
~~\vline &\vline & \vline &\vline~~ 
\end{bmatrix} ^T.
\end{eqnarray*}
As an illustration, for polynomials (with $U=0$) we would have 
\begin{eqnarray*}
\Phi(Y, U) = 
\begin{bmatrix} 
{1}& Y & Y^{P_2}  & \cdots ~~
\end{bmatrix}
\end{eqnarray*}
where higher polynomials are denoted as $Y^{P_2}, Y^{P_3},$ etc. For instance, $Y^{P_2}$ denotes quadratic nonlinearities:
\begin{eqnarray*}
Y^{P_2} = \begin{bmatrix}
y_1^2(1) & y_1(1)y_2(1) & \cdots & y_n^2(1)\\
y_1^2(2) & y_1(2)y_2(2) & \cdots   & y_n^2(2)\\
\vdots &\vdots &\ddots& \vdots\\
y_1^2(M) & y_1(M)y_2(M) & \cdots  & y_n^2(M)
\end{bmatrix}.
\end{eqnarray*}
Basically, each column of $\Phi(Y, U)$ represents a candidate function for a nonlinearity in $\bf{f}_k$. \\ \\
These libraries of functions may be very large. However, since only a very small number of these nonlinearities appear in each row of $\Phi$, we set up a sparse regression problem to determine the sparse vectors of coefficients $W = \begin{bmatrix}w_1 &w_2 & \ldots & w_n\end{bmatrix}$. The nonzero elements in $W$ determine which nonlinearities are active and the corresponding parameters. Letting
\begin{eqnarray*}
\bar{Y}\triangleq
\begin{bmatrix}
y_1(2)  & \cdots & y_1(M+1)\\
y_2(2)  & \cdots & y_2(M+1)\\
\vdots  & \ddots & \vdots \\
y_n(2)  & \cdots & y_n(M+1)
\end{bmatrix}^T,
\end{eqnarray*}
and define residual $Z = \bar{Y} -\Phi W -\noise$, then the first objective is to find the sparsest possible $Z$, i.e., 
\begin{equation*}
\begin{aligned}
W^*&=\arg\min_W \|Z\|_{\ell_0}\\
& \text{subject to:}~Z = \bar{Y} -\Phi W -\noise.
\end{aligned}
\end{equation*} 
This step identifies which time points correspond to which subsystem. The second objective performs a similar optimization, but only for those data points where $Z$ is zero, and searching for sparse $W$. This step identifies the actual dynamics of each subsystem. These two steps are done iteratively until all subsystems have been identified. Further details are found in Algorithm~1 in Supplementary Materials. 

\subsubsection*{Transition logic identification}
Define $\gamma_{i}(t)$ as the set membership: it equals to $1$ only if the subsystem $i$ is active at discrete-time $t$, otherwise it equals to $0$. These functions are known from the information in the subsystem identification above. Here, we are interested in learning what functions trigger the switch from one subsystem to another. Define also $\text{step}(x)$, which equals to $1$ if $x\ge0$, and $0$ otherwise. 
Mathematically, we are searching for a nonlinear function $g$, such that $\text{step}(g(y(t),u(t)))$  specifies the membership. Due to non-differentiability of step functions at $0$, we alternatively relax the step function to a sigmoid function, i.e., $\gamma_{j}(t+1)  \approx \frac{1}{1+e^{-g(y(t),u(t))}}$, where $j$ is a potential subsystem that we can jump to at time $t+1$. Assuming we are in subsystem $i$ at time $t$, the fitness function to jump to subsystem $j$ at time $t+1$ is then
\begin{equation}\label{eq:tm11}
\begin{aligned}
\sum_{t=1}^{M-1} \gamma_i (t)\left \| \gamma_{j}(t+1) - \frac{1}{1+e^{-g(y(t),u(t))}} \right\|_{\ell_2}^2.
\end{aligned}
\end{equation}
To minimize eq.~\eqref{eq:tm11}, we can parameterize $g(y(t),u(t))$ as a linear combination of over-determined dictionary matrix, i.e.,  $g(y(t),u(t)) \triangleq \Psi(Y, U) v$, in which $\Psi$ can be constructed similarly to $\Phi$ in the previous subsection and $v$ is a vector of to-be-discovered parameters. Further details are found in Algorithm~2 in Supplementary Materials. 

\section*{Acknowledgements}
{\bf General:} The first author would like to sincerely thank Prof. Claire J. Tomlin for insightful discussion and continuous help. We thank Prof. Guang Yang for help on the experimental setup. We thank Mr. Anthony Haynes, Mr. Frank Jiang, Dr. Anija Dokter and Mrs. Karen Haynes for editing. We thank Dr. Daniel Ly and Prof. Ke Li for sharing datasets. {\bf Author contributions:} Y.Y. developed the IHYDE algorithms. Y.Y. and X.T. developed simulation codes for the example problems considered. All authors participated in designing and discussing the study and writing the paper. {\bf Competing interests:} The authors declare that they have no competing interests. {\bf Data and materials availability:} All data needed to evaluate the conclusions in the paper are present in the paper and/or the Supplementary Materials. The code implementation of IHYDE is available at https://github.com/HAIRLAB/CPSid.

\clearpage
\phantom{s}
 \thispagestyle{empty}
\clearpage



\section*{Supplementary material}

# Supplementary Materials

# Contents

# S1 Preliminaries

## S1.1 Notations

$\mathbb{R}^n$ denotes the n-dimensional Euclidean space.

$\mathbb{Z}$ denotes the set of integers, $\ldots, -1, 0, 1, \ldots$.

$\|x\|_{\ell_0}$: the $\ell_0$-norm of a vector $x$.

$\|x\|_{\ell_1}$: the $\ell_1$-norm of a vector $x$, i.e., $\|x\|_{\ell_1} = \sum_{i=1}^{n} |x_i|$.

$\|x\|_{\ell_2}$: the $\ell_2$-norm of a vector $x$, i.e., $\|x\|_{\ell_2} = (|x_1|^2 + \cdots + |x_n|^2)^{1/2}$.

$A$: For a matrix $A \in \mathbb{R}^{M \times N}$, $A[i,j] \in \mathbb{R}$ denotes the element in the $i^{th}$ row and $j^{th}$ column, $A[i,:] \in \mathbb{R}^{1 \times N}$ denotes its $i^{th}$ row, $A[:,j] \in \mathbb{R}^{M \times 1}$ denotes its $j^{th}$ column.

$\alpha$: For a column vector $\alpha \in \mathbb{R}^{N \times 1}$, $\alpha[i]$ denotes its $i^{th}$ element.

## S1.2 Introduction to Hybrid Dynamical Systems

Before giving formal definition to hybrid dynamical systems, we shall first give a brief review to dynamical systems. A dynamical system describes how state variables (typically physical quantities) evolve with respect to time. Following definitions in *[36]*, we define three types of variables.

1. continuous state variables: if the state variable takes value in $\mathbb{R}^n$ for $n \geq 1$.

2. discrete state variables: if the state variable takes value in a finite set, for example, $\{1, 2, 3, \ldots\}$.

3. hybrid state variables: if a part of the state variables are continuous and the other discrete.

Based on the time set over which the state evolves, we classify the dynamical systems as:

- continuous time: if the set of time is a subset of the real line $\mathbb{R}$. Normally we use $t \in \mathbb{R}$ to denote the continuous time. The evolution of the state-variables in continuous time can be described as ordinary differential equations.

- discrete time: if the set of time is a subset of the integers. Normally we use $k \in \mathbb{Z}$ to denote discrete time. The evolution of the state-variables in discrete time can be described as difference equations.

A hybrid dynamical system $\mathcal{H}$, is defined as a tuple, $\mathcal{H} = (\mathcal{W}, \mathcal{M}, \mathcal{F}, \mathcal{T})$ with the following definitions:

- $\mathscr{W}$ defines a subspace in $\mathbb{R}^{m+n}$ for input-output variables $u(t) \in \mathbb{R}^m, y(t) \in \mathbb{R}^n$;

- $\mathscr{M}$ defines a countable, discrete set of modes in which only a single mode, $m \in \{1, 2, \ldots, K\}$, is occupied at a given time;

- $\mathscr{F}$ defines a countable discrete set of first-order differential equations:

$$\mathscr{F} = \left\{ \frac{dy(t)}{dt} = \mathbf{F}_k(y(t), u(t))) \mid k = 1, 2, \ldots, K \right\}$$

- $\mathscr{T}$ defines a countable discrete set of transitions, where $\mathscr{T}_{i \to j}$ denotes a Boolean expression that represents the condition to transfer from mode $i$ to $j$.

The signals $y(t)$ and $u(t)$ are sampled at a rate $h > 0$, i.e. sampled at times $0, h, 2h, 3h....$ For fast enough sampling (or low $h$), standard system identification typically obtains first a discrete-time system, and then coverts it to a continuous-time system [28]. One of the simplest methods to approximate derivatives is to consider

$$\frac{dy(t)}{dt} \approx \frac{y(t+h) - y(t)}{h}$$

which yields the discrete-time system

$$y(t+h) = y(t) + h\, \mathbf{F}_k(y(t), u(t)) \triangleq \mathbf{f}_k(y(t), u(t)), \quad k \in \{1, 2, \ldots, K\}$$

For simplification of notation, assume the system can be written as

$$y(t+h) = \mathbf{f}_k(y(t), u(t)) \triangleq \mathbf{I}_k(y(t)) + \mathbf{h}_k(u(t)), \quad k \in \{1, 2, \ldots, K\}$$

Hence, the class of systems considered is discrete-time, Markovian and nonlinear. While this is already a very rich class of systems, it can be easily extended to more general nonlinear systems, including, for example, dynamics of non-separable nonlinear functions of $(y(t), u(t))$.

Without loss of generality, we can rescale the time variable $t$ so that $h = 1$. Thus, we can construct a mathematical model for hybrid dynamical systems

$$\begin{aligned} m(t+1) &= \mathscr{T}(m(t), y(t), u(t)) \\ y(t+1) &= \mathbf{f}(m(t), y(t), u(t)) = \begin{cases} \mathbf{f}_1(y(t), u(t)), & \text{if } m(t) = 1 \\ \quad \vdots, & \vdots \\ \mathbf{f}_K(y(t), u(t)), & \text{if } m(t) = K \end{cases} \end{aligned} \quad (2)$$

**Example 1.** *Consider again the temperature control system in Fig. 1, consisting of a heater and a thermostat. Variables that are included in a model of such a system are the room temperature y(t) and the operating mode of the heater (on or off). Assuming a sampling time of h > 0, we obtain the following approximate difference equations for the temperature*

$$\text{Subsystem 1 (heat off)}: \frac{y(t+1) - y(t)}{h} \approx -ay(t), \Rightarrow y(t+1) = (1-ah)y(t)$$

$$\text{Subsystem 2 (heat on)}: \frac{y(t+1) - y(t)}{h} \approx -a(y(t) - 30), \Rightarrow y(t+1) = (1-ah)y(t) + 30ah.$$

*These two equations model how the temperature changes under the heater off or on, respectively. The transition logics between the two subsystems are*

$$\text{Transition logic from subsystem 1 to 2 } \mathcal{T}_{1 \to 2}: y \leq 19$$

$$\text{Transition logic from subsystem 2 to 1 } \mathcal{T}_{2 \to 1}: y \geq 21,$$

*representing the control for the operating mode of the heater. This is the hybrid dynamical system model for the thermostat. Given the hybrid dynamical system, we can study the stability of such a system or simulate the system to check the trajectories of state variables. Note that, in practice, the hybrid dynamical system model is usually unknown or only partially known. The goal of this paper is to infer both the above subsystems and the transition logic (Fig. 1(A)) from only time-series data of the temperature in Fig. 1(C).*

## S2  IHYDE Algorithm

When a hybrid dynamical system has a single subsystem, i.e., $K = 1$ in eq. (2), it becomes a time-invariant nonlinear dynamical system. We start by briefly reviewing identification tools for this class of systems from *[10, 13]*, since parts of our proposed algorithm are based on these tools. As explained before, our algorithm uses only time-series data to directly model the system. Hence, the first step is to collect time-course input-output data $(y(t), u(t))$ uniformly sampled at a number of discrete time indices $t = 1, 2, \ldots, M+1$. Let

$$Y = \begin{bmatrix} | & | & & | \\ y(1) & y(2) & \cdots & y(M) \\ | & | & & | \end{bmatrix}^T, \quad U = \begin{bmatrix} | & | & & | \\ u(1) & u(2) & \cdots & u(M) \\ | & | & & | \end{bmatrix}^T.$$

Note that $y(t) \in \mathbb{R}^n$ and $u(t) \in \mathbb{R}^m$, and so $Y \in \mathbb{R}^{M \times n}$ and $U \in \mathbb{R}^{M \times m}$. Next, we construct an overdetermined library $\Phi(Y, U)$ consisting of potential nonlinear functions that appear in $\mathbf{f}_k$ in eq. (2). It is

expected that the true nonlinearities are part of this library. The choice of these functions is guided by the particular field of study. For example, the library would consist of sinusoidal functions in pendulums, and polynomial and sigmoidal functions in biochemical networks. As an illustration, a library consisting of constant or polynomials would result in the following dictionary matrix

$$\Phi(Y,U) = \begin{bmatrix} 1 & Y & Y^{P_2} & \cdots & U & U^{P_2} & \cdots \end{bmatrix}.$$

Here, higher polynomials are denoted as $Y^{P_2}, Y^{P_3}$, etc. For example, $Y^{P_2}$ denotes the quadratic nonlinearities in the state variable $Y$, given by:

$$Y^{P_2} = \begin{bmatrix} y_1^2(1) & y_1(1)y_2(1) & \cdots & y_n^2(1) \\ y_1^2(2) & y_1(2)y_2(2) & \cdots & y_n^2(2) \\ \vdots & \vdots & \ddots & \vdots \\ y_1^2(M) & y_1(M)y_2(M) & \cdots & y_n^2(M) \end{bmatrix}.$$

Basically, each column of $\Phi(Y,U)$ represents a candidate function for a nonlinearity in **f**. The library of functions may be very large. However, since only a very small number of these nonlinearities appear in each row of $\Phi(Y,U)$, we can set up a sparse regression problem to determine the sparse matrices of coefficients $W = \begin{bmatrix} w_1 & w_2 & \cdots & w_n \end{bmatrix}$, where $w_i \in \mathbb{R}^{P \times 1}$ and $P$ is the total number of total number of candidate functions in the library. The nonzero elements in $W$ determine which nonlinearities are active *[10, 13]* and the corresponding parameters.

Let

$$\bar{Y} \triangleq \begin{bmatrix} y_1(2) & \cdots & y_n(2) \\ y_1(3) & \cdots & y_n(3) \\ \vdots & \ddots & \vdots \\ y_1(M+1) & \cdots & y_n(M+1) \end{bmatrix}.$$

This results in the overall model $\bar{Y} = \Phi(Y,U)W + \xi$, where $\xi = \begin{bmatrix} \xi_1 & \xi_2 & \cdots & \xi_n \end{bmatrix}$ and $\xi_i \in \mathbb{R}^{M \times 1}$ is zero-mean i.i.d. Gaussian measurement noise with covariance matrix $\sigma^2 I$, for some $\sigma \geq 0$. The work in *[10, 13]*, developed methods based on Lasso and Sparse Bayesian Learning for identifying each $w_i$ in the above equation as the following optimization:

$$w_i^* = \arg\min_{w} \|\bar{Y}[:,i] - \Phi w_i\|_{\ell_2}^2 + \lambda \|w_i\|_{\ell_1}. \tag{3}$$

For the completeness of this paper, we briefly review the results in *[10]* in the Appendix A.

## S2.1 Inferring Sub-systems

When $K > 1$, we can use a similar formulation as above. However, the challenge is that there is no single $W$ typically fits all the data due to the hybrid nature of the dynamical system. Next, we introduce a new method to tackle such a challenge.

Define $Z = \bar{Y} - \Phi W - \boldsymbol{\xi}$, where $\boldsymbol{\xi}$ is defined similarly as realizations of zero-mean i.i.d. Gaussian measurement noise with covariance matrix $\sigma^2 I$. The goal is to find a $Z^* \triangleq \begin{bmatrix} Z_1^* & Z_2^* & \ldots & Z_n^* \end{bmatrix} \triangleq \bar{Y} - \Phi W^* - \boldsymbol{\xi}$ as sparse as possible, i.e.,

$$W^* = \arg\min_W \sum_{i=1}^n \|Z_i\|_{\ell_0} \tag{4}$$
$$\text{subject to: } Z = \bar{Y} - \Phi W - \boldsymbol{\xi}.$$

Correspondingly, we have $Z_i^* = \bar{Y} - \Phi w_i^*$. The interpretation of this optimization is to find a $W$ (or equivalently a subsystem) that fits most of the input-output data. As a result, the index of the zero entries of $Z^*$ corresponds to the index for input-output that is can be fitted by a single subsystem. This initial idea was similar to those presented in *[18]*, yet we later extend this idea to a robust Bayesian algorithm that works well for noisy data. To solve eq. (4), assume, without loss of generality, that the dictionary matrix $\Phi$ is full rank. Define a transformation matrix $\Theta$ in which each column spans the left null space of $\Phi$. Then, it follows that $\Theta\bar{Y} = \Theta Z + \Theta\boldsymbol{\xi}$. Using an appropriate Lagrange multiplier $\lambda_z$, we now can rewrite the above problem as an unconstrained minimization:

$$\min_Z \frac{1}{2}(\tilde{\bar{Y}} - \Theta Z)^T \Sigma^{-1} (\tilde{\bar{Y}} - \Theta Z) + \lambda_z \sum_{i=1}^n \|Z_i\|_{\ell_0},$$

where $\tilde{\bar{Y}} \triangleq \Theta\bar{Y}$ and $\Sigma = \Theta\Theta^T$.

**Remark 1.** *This is the key step in the later proposed algorithm; there is no W in this optimization after the transformation. Instead, we are optimizing over the residual Z.*

However, this problem is known to be computationally expensive. Instead, we use the following convex relaxation

$$Z^* = \arg\min_Z \frac{1}{2}(\tilde{\bar{Y}} - \Theta Z)^\top \Sigma^{-1} (\tilde{\bar{Y}} - \Theta Z) + \lambda_z \sum_{i=1}^n \|Z_i\|_{\ell_1}.$$

We can decompose the above optimization to a number of smaller optimizations: for $i = 1, \ldots, n$

$$Z_i^* = \arg\min_{Z_i} \frac{1}{2}(\tilde{\bar{Y}}[:,i] - \Theta Z_i)^\top \Sigma^{-1} (\tilde{\bar{Y}} - \Theta Z_i) + \lambda_z \|Z_i\|_{\ell_1}. \tag{5}$$

Once this problem is solved, we consider the index set seq $= \{j | |Z_i^*[j]| \leq \varepsilon_z\}$ and further identify the sparse coefficients $w_i^*$ using the following optimization

$$w_i^* = \arg\min_{w_i} \frac{1}{2}\|\bar{Y}[\text{seq},i] - \Phi[\text{seq},:]w_i\|_{\ell_2}^2 + \lambda_w \|w_i\|_{\ell_1}.$$

The variables $w_i^*$ are the coefficients of the identified subsystem.

**Remark 2.** *The reason to enforce $w_i^*$ to be sparse is due to the constructed redundant dictionary matrix $\Phi$.*

We further define error $= abs(\bar{Y}[:,i] - \Phi w_i^*)$ (here *abs* is an elementary-wise operator which returns the absolute value of every element of a vector) and we set the $j$th element of $\bar{Y}[:,i]$: $\bar{Y}[j,i] = 0$ and the $j$th row of $\Theta$: $\Theta[j,:] = 0$ if the $j$th element of error is less than $\varepsilon_w$, for some small $\varepsilon_w > 0$. This removes the data that has already been fitted by the subsystem.

Once we have the new $\bar{Y}$ and $\Theta$, we can solve the same problem with the remaining time points (where the corresponding elements of $\bar{Y}$ and the corresponding row of $\Theta$ are nonzero) using the exact same procedure. The number of iterations gives the minimum number of subsystems. The proposed algorithm is summarized in Algorithm 1. The code implementation is available at https://github.com/HAIRLAB/CPSid with User's Manual in Appendix B. In what follows, we shall briefly discuss extensions and variants of Algorithm 1, which can empirically improve the performance of IHYDE.

### S2.1.1 Application to nonhybrid dynamical systems

When there is only one subsystem, we show that $Z^i$ should be an all 0 matrix from the first optimization in eq. (5). Eq. (6) should be the same as eq. (3) since seq $= \{1,2,\ldots,M+1\}$, which recovers the results for time-invariant nonlinear system identification in *[10, 13]*. As a result, IHYDE provides a unified point of view to the subsystems identification problem for any $K \in \{1,2,\ldots\}$.

### S2.1.2 Variants of Algorithm 1

One of the issues in Algorithm 1, according to the experiments presented in the Results section, is quite sensitive to both noise and redundancy in the dictionary matrix. The reason for this sensitivity is that the proposed algorithm has a sequence of optimizations, thus, when an error occurs early on, the optimization propagates it to later times, causing large final errors. We hereby propose a new variant of Algorithm 1 to improve the empirical performance of optimizations using Bayesian statistics. The derivations are presented in Appendix A.

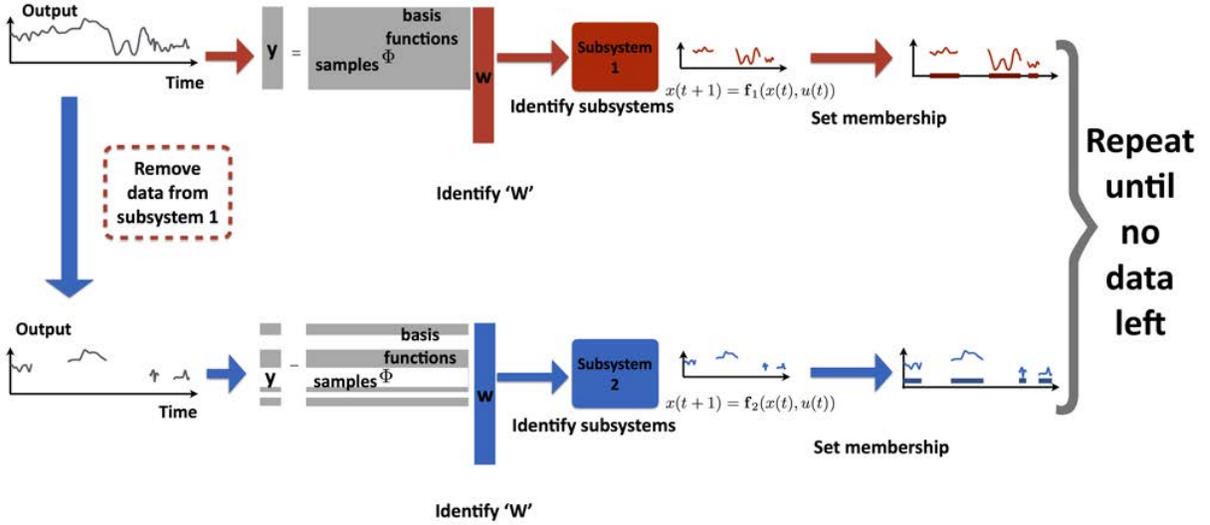

**Fig. S1. Schematics of the proposed subsystems identification algorithm.** We construct a library of nonlinear functions to start with and a dictionary matrix $\Phi$. We formulate an iterative convex optimization method to infer the number of subsystems and the underlying system models for every subsystem. More specifically, we first identify a best model that fits the majority of data, then we remove the fitted data and re-do the identification until no data are left.

## S2.2 Inferring Transition Logic

Once the subsystems have been identified, we can assign every input-output data point $(u(t), y(t))$ to a specific subsystem as shown in Fig. S1. The next step is to identify the transition logic between different subsystems. We first convert the problem of identifying the transition logic to a standard sparse logistic regression problem which can be efficiently solved by many methods in the literature. The scheme is illustrated in Fig. S2.

To proceed, we define $\gamma_i(t)$ as the set membership which equals to 1 only if the subsystem $i$ is active at discrete-time $t$ or otherwise it equals to 0. The goal is to identify the transition rules $\mathcal{T}_{i \to j}$ between any subsystems $i, j$. These functions are known from the information in the subsystem identification above. Define also step$(x)$, which equals 1 if $x \geq 0$, and 0 otherwise. Mathematically, we are searching for a nonlinear function $g$, such that step$(g(y(t), u(t)))$ specifies the membership. Due to non-differentiability of step functions at 0, we alternatively relax the step function to a sigmoid function, i.e., $\gamma_j(t+1) \approx \frac{1}{1+e^{-g(y(t),u(t))}}$, where $j$ is a potential subsystem that we can jump to at time $t+1$. Assuming we are in subsystem $i$ at time $t$, the fitness function to jump to subsystem

**Algorithm 1** Sub-systems Identification Algorithm
───────────────────────────────────────────────
1: **Input:** Collect input-output data $u(t)$ and $y(t)$ for $t = 1, 2, \ldots, M+1$. Two pre-specified thresholds $\varepsilon_z$ and $\varepsilon_w$, two tuning parameters $\lambda_z$ and $\lambda_w$
2: **Output:** Return $\{W^i\}$ for $i = 1, \ldots, K$ and the number of subsystems $K$
3: Construct dictionary matrix $\Phi(Y, U)$ based on prior knowledge of the system
4: **for** $j = 1, \ldots, n$ **do**
5:     **for** $i = 1, \ldots, K_{max}$ **do**
6:         Compute $\Theta$ in which all column span the left null space of $\Theta$: $\Theta\Phi = 0$
7:         Solve for $Z_j^i$ from eq. (5)
8:         **if** $Z_j^i = 0$ **then**
9:             $K = i$, Break
10:        **end if**
11:        $h = 1$ and seq $= [\ ]$
12:        **for** $l = 1 \ldots, M$ **do**
13:            **if** the $l$th element of $Z_j^i$, i.e., $abs(Z_j^i[l]) \leq \varepsilon_z$ **then**
14:               Set seq$[h] = l$ and $h{+}{+}$
15:            **end if**
16:        **end for**
17:        Solve the following convex optimization

$$w_j^i = \arg\min_w \frac{1}{2}\|\bar{Y}[\text{seq}, j] - \Phi[\text{seq}, :]w\|_{\ell_2}^2 + \lambda_w \|w\|_{\ell_1} \quad (6)$$

18:        error $= abs(\bar{Y}[:, j] - \Phi w_j^i)$
19:        **for** $l = 1 \ldots, M$ **do**
20:            **if** the $l$th element of error, i.e., error$[l] \leq \varepsilon_w$ **then**
21:               Set $\bar{Y}[l, j] = 0$ and $\Theta[l, :] = 0$
22:            **end if**
23:        **end for**
24:     **end for**
25: **end for**
26: Return nonzero $W^i \triangleq \begin{bmatrix} w_1^i & \ldots & w_n^i \end{bmatrix}$ for $i = 1, \ldots, K$ and the number of subsystems $K$
───────────────────────────────────────────────

$j$ at time $t+1$ is then

$$\sum_{t=1}^{M-1} \gamma_i(t) \left\| \gamma_j(t+1) - \frac{1}{1 + e^{-g(y(t), u(t))}} \right\|_{\ell_2}^2. \quad (7)$$

To solve the optimization in (7), we can parameterize $g(y(t), u(t))$ as a linear combination of over-determined dictionary matrix, i.e., $g(y(t), u(t)) \triangleq \Psi(y(t), u(t))v$, in which $\Psi$ can be constructed similarly as $\Phi$ in the previous section and $v$ is a vector of to-be-discovered parameters. The cost function only takes non-zero value when $\gamma_i(t) = 1$.

28

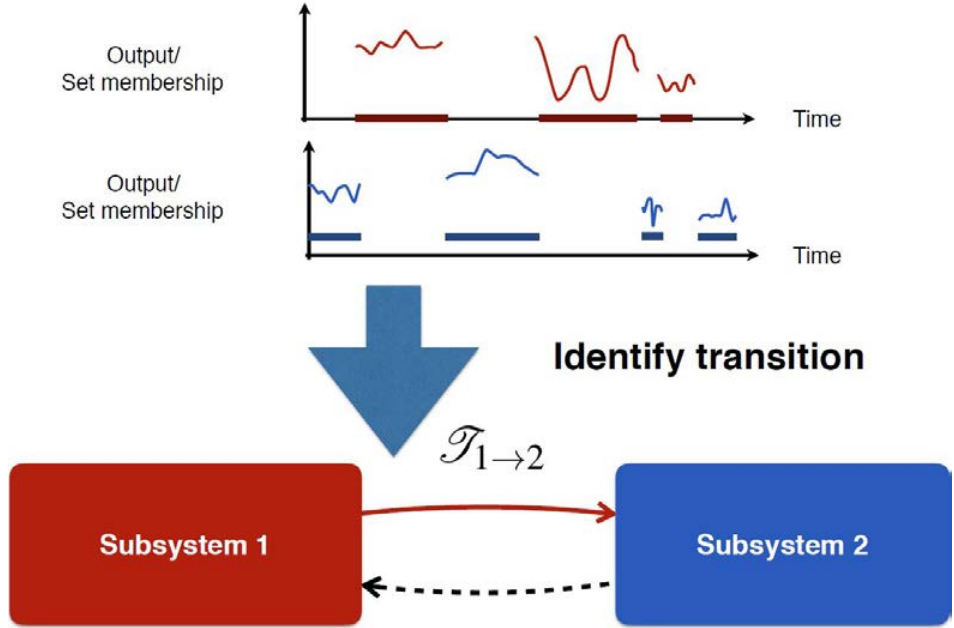

**Fig. S2. Illustration of the proposed Algorithm to identify transition logic.** Using the membership of every classified data point, we apply logistic regression to infer the logic between every pair of identified subsystems, i.e., $\mathcal{T}_{i \to i'}$ for every $i$ and $i'$.

Let $V \triangleq \{t | \gamma_i(t) = 1\}$, then

$$\sum_{t=1}^{M-1} \gamma_i(t) \left\| \gamma_j(t+1) - \frac{1}{1+e^{-g(y(t),u(t))}} \right\|_{\ell_2}^2 = \sum_{k \in V} \left\| \gamma_j(t+1) - \frac{1}{1+e^{-\Psi v}} \right\|_{\ell_2}^2. \tag{8}$$

After this transformation, the minimization of eq. (8) is known as a the logistic regression. Hence, we can use the standard gradient descent method to solve the logistic regression *[39]*.

Similarly, we can also add an $\ell_1$ regularizer in the optimization, i.e., we minimize

$$\sum_{t \in V} \left\| \gamma_j(t+1) - \frac{1}{1+e^{-\Psi v}} \right\|_{\ell_2}^2 + \beta \|v\|_{\ell_1}. \tag{9}$$

There are many Matlab codes for sparse linear logistic regression. Here, we adopt the implementation framework proposed in *[44]*.

## S2.3 Cross Validation for Parameter Tuning

We use cross-validation as a model validation technique for assessing the results of the identified system in our experiments. We firstly identify a model from a dataset, and later we test it on a dataset which is unseen in the modeling stage. The goal of cross validation is to prevent prob-

**Algorithm 2** Transition Logic Identification Algorithm
---
1: **Input:** Input-output data $y(t), u(t)$ and $\gamma_i(t)$, $i = 1, 2\ldots, K$ and $t = 1, 2, \ldots, M$
2: **Output:** Transition logic $\mathcal{T}_{i \to j}(y(t), u(t))$ for any pair $i, j$
3: **for** $i = 1, \ldots, K$ **do**
4:     **for** $j \neq i$ **do**
5:         Construct the dictionary matrix from prior knowledge $\Psi$ as described in the main text
6:         The solution to the logistic regression in eq. (9) gives the transition model for $\mathcal{T}_{i \to j}$
7:     **end for**
8: **end for**
9: Return all transition logic mapping $\mathcal{T}$
---

lems like overfitting, give an insight on how the model can be generalized to different datasets. In particular, we use $s$-fold cross-validation. The original sample is randomly partitioned into $s$ equal sized subsamples. Of the $s$ subsamples, a single subsample is retained as the validation data for testing the model, and the remaining $s-1$ subsamples are used as training data. The cross-validation process is then repeated $s$ times, with each of the $s$ subsamples used exactly once as the validation data. The $s$ results from the folds can then be averaged to produce a single estimation. The advantage of this method over repeated random sub-sampling is that all observations are used for both training and validation, and each observation is used for validation exactly once.

# S3  Results for IHYDE

This section applies IHYDE to a number of examples ranging from power systems to robotics, showcasing the wide range of applicability of the proposed IHYDE method.

## S3.1  Example 1: Hysteresis Relay

One of the most common Cyber Physical Systems is the Hysteresis Relay. It is found, for example, in almost all thermostats: the heater is turned on when the temperature is below a threshold, and turned off when the temperature is above another threshold. Typically, the low and high temperature switching are different to avoid frequent switching, which could damage the system. The Hysteresis Relay can be found in physical, chemical, engineering and biological applications.

The datasets for discovery are generated by Ly et. al. in *[35]*. The additive noise level varies from 0% to 6% in 2% increments, i.e., $N_p = \frac{\sigma_{noise}}{\sigma_y} \times 100\%$, where $\sigma_{noise}$ is the noise variance and $\sigma_y$ is the variance of the measurement. We apply the proposed IHYDE to data generated by an unknown Hysteresis Relay to discover its hybrid dynamical model (shown in Fig. S3). The discovered systems are shown in Table S1 and Table S2 using 2000 data-points repsectively.

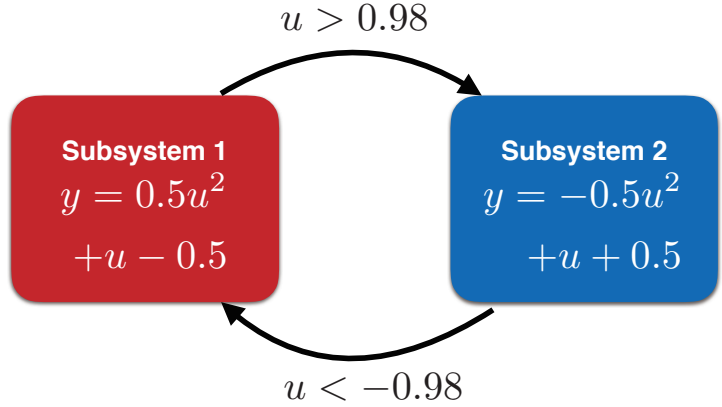

**Fig. S3.** The hybrid dynamical system model for the Hysteresis Relay. *u* and *y* are the measured outputs of the Hysteresis Relay.

Using IHYDE for subsystems identification, we are able to successfully identify that there are only two subsystems that generate the datasets. In addition, the two identified subsystems are consistent with or close to the true ones from both noiseless and noisy data. Specifically, with or without redundant basis functions, we are able to identify the true systems, achieving very similar discovery results. This, in other words, demonstrates that the IHYDE is able to discover the true subsystems, the number of subsystems together with parameterizations of every subsystem.

**Table S1.** The identified systems of Hysteresis Relay for both noiseless and noisy datasets. We also present the tuning parameters for different noise levels.

| Noise | $N_p = 0$ | $N_p = 2\%$ | $N_p = 4\%$ | $N_p = 6\%$ |
|---|---|---|---|---|
| Library $\Phi$ | 1 | | | |
| Identified subsystem 1 | $y = 0.9999$ | $y = 1.0013$ | $y = 1.0027$ | $y = 1.0040$ |
| Identified subsystem 2 | $y = -0.9999$ | $y = -1.0004$ | $y = -1.0009$ | $y = -1.0014$ |
| $\lambda_z$ | $1e-3$ | | | |
| $\varepsilon_z$ | $1e-4$ | | | |
| $\lambda_w$ | 0.05 | | | |
| $\varepsilon_w$ | | 0.2 | | 0.3 |
| Number of misclassified points | 0 | | | |

**Table S2.** The identified result and details of Hysteresis Relay with redundant basis functions.

| Noise | $N_p = 0$ | $N_p = 2\%$ | $N_p = 4\%$ | $N_p = 6\%$ |
|---|---|---|---|---|
| Library $\Phi$ | $1\ u\ u^2\ u^3\ u^4\ u^5$ | | | |
| Identified subsystem 1 | $y = 0.9999$ | $y = 1.0013$ | $y = 1.0027$ | $y = 1.0038$ |
| Identified subsystem 2 | $y = -0.9999$ | $y = -1.0004$ | $y = -1.0009$ | $y = -1.0014$ |
| $\lambda_z$ | $1e-3$ | | | |
| $\varepsilon_z$ | $1e-4$ | | | |
| $\lambda_w$ | 0.05 | | | |
| $\varepsilon_w$ | 0.2 | | | |
| Number of misclassified points | 0 | | | |

Once all subsystems have been identified and all data points have been classified, IHYDE identifies the transition logic between subsystems. Even when there exists redundant basis function, IHYDE is able to precisely identify the correct transition logic. The identified results are shown in Table S3 and Table S4.

**Table S3.** The identified transition logic of the Hysteresis Relay using noiseless data and redundant basis functions.

| Systems | Subsystem 1 | Subsystem 2 |
|---|---|---|
| Subsystem 1 | | $u > 0.4995$ |
| Subsystem 2 | $u < -0.4987$ | |
| Library $\Psi$ | 1 | $u$ |
| $\beta$ | 0.1 | |

**Table S4.** The identified transition logic of the Hysteresis Relay using noiseless data and redundant basis functions.

| Systems | Subsystem 1 | Subsystem 2 |
|---|---|---|
| Subsystem 1 | | $u > 0.4995$ |
| Subsystem 2 | $u < -0.4987$ | |
| Library $\Psi$ | 1 $\quad u \quad e^{(10*(\sin(u^2))+10)} \quad \frac{\log(|u|)}{\sin(u)} \quad u^2 \quad u^4$ | |
| $\beta$ | 0.1 | |

## S3.2 Example 2: Continuous Hysteresis Loop

A Continuous Hysteresis Loop is yet another classical hybrid system– here we use the Preisach model *[35]* for data simulation. In our setup, each subsystem has its own input-output behavior while the transitions occur when the input hits certain thresholds. We apply the IHYDE to reverse engineering the Continuous Hysteresis Loop using 2000 data points generated by *[35]*.

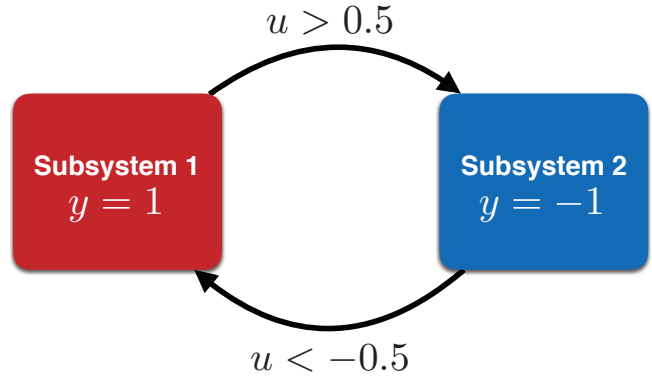

**Fig. S4.** The hybrid dynamical systems model for the Continuous Hysteresis Loop.

The identified systems are shown in Table S5 and Table S6. In contrast with the previous Hysteresis Relay example, the IHYDE starts obtaining false classification results as the noise level increases and with redundant basis functions. Yet, IHYDE is still able to identify the actual subsystem dynamics up to some precision.

Once all subsystems have been identified and all data points have been classified, IHYDE identifies the transition logic between subsystems. When there does not exist redundant basis function (i.e., when prior knowledge is available about the structure of transition logic), IHYDE is able to precisely identify the correct transition logic. The identified results are shown in Table S7. When there exists redundant basis functions, IHYDE still successfully identifies the transition logic. The identified results are shown in Table S8.

**Table S5.** The identified result and details of Continuous Hysteresis Loop.

| Noise | $N_p = 0$ | $N_p = 2\%$ | $N_p = 4\%$ | $N_p = 6\%$ |
|---|---|---|---|---|
| Library $\Phi$ | polynomials in $u$ up to second order | | | |
| Identified subsystem 1 | $y = 0.4998u^2$ $+1.0000u$ $-0.4999$ | $y = 0.4989u^2$ $+1.0003u$ $-0.4996$ | $y = 0.4920u^2$ $+0.9999u$ $-0.4982$ | $y = 0.4842u^2$ $+0.9980u$ $-0.4961$ |
| Identified subsystem 2 | $y = -0.5042u^2$ $+1.0021u$ $+0.5008$ | $y = -0.5072u^2$ $+1.0031u$ $+0.5015$ | $y = -0.5172u^2$ $+1.0055u$ $+0.5032$ | $y = -0.5133u^2$ $+0.9966u$ $+0.5027$ |
| $\lambda_z$ | 0.005 | 0.005 | 0.008 | 0.03 |
| $\varepsilon_z$ | $1e-4$ | | | |
| $\lambda_w$ | 0.005 | 0.001 | 0.005 | 0.008 |
| $\varepsilon_w$ | 0.04 | 0.04 | 0.08 | 0.105 |
| Number of misclassified points | 0 | 17 | 45 | 88 |

**Table S6.** The identified result and details of Continuous Hysteresis Loop with redundant basis functions.

| Noise | $N_p = 0$ | $N_p = 2\%$ | $N_p = 4\%$ | $N_p = 6\%$ |
|---|---|---|---|---|
| Library $\Phi$ | $1\ u\ u^2\ e^{-u}\ \frac{u^3}{e^u}\ \frac{\cos(2u)}{\sin(u)^3}$ | | | |
| Identified subsystem 1 | $y = 0.4999u^2$ $+1.0000u$ $-0.4999$ | $y = 0.5001u^2$ $+1.0001u$ $-0.4998$ | $y = 0.4919u^2$ $+0.9995u$ $-0.4979$ | $y = 0.4811u^2$ $+0.9994u$ $-0.4956$ |
| Identified subsystem 2 | $y = -0.5010u^2$ $+0.9995u$ $+0.5002$ | $y = -0.4979u^2$ $+0.9990u$ $+0.5001$ | $y = -0.5123u^2$ $+1.0000u$ $+0.5034$ | $y = -0.5275u^2$ $+0.9999u$ $+0.5047$ |
| $\lambda_z$ | 0.005 | 0.005 | 0.008 | 0.03 |
| $\varepsilon_z$ | $1e-4$ | | | |
| $\lambda_w$ | 0.005 | 0.001 | 0.005 | 0.008 |
| $\varepsilon_w$ | 0.04 | 0.04 | 0.08 | 0.105 |
| Number of misclassified points | 0 | 11 | 45 | 94 |

**Table S7.** The identified transition logic of Continuous Hysteresis Loop.

| System | Subsystem 1 | Subsystem 2 |
|---|---|---|
| Subsystem 1 | | $u > 0.9803$ |
| Subsystem 2 | $u < -0.9799$ | |
| Library $\Psi$ | 1 | $u$ |
| $\beta$ | 10 | |

**Table S8.** The identified transition logic of Continuous Hysteresis Loop when existing redundant basis functions.

| System | Subsystem 1 | Subsystem 2 | | |
|---|---|---|---|---|
| Subsystem 1 | | $u > 0.9803$ | | |
| Subsystem 2 | $u < -0.9799$ | | | |
| Library $\Psi$ | 1 | $u$ | $\frac{1}{u^2}$ | $\frac{\cos u}{\sin u^3}$ |
| $\beta$ | 10 | | | |

## S3.3 Example 3: Phototaxis Robot

Consider a Phototaxis Robot with a hybrid dynamical system model shown in Fig. S5 *[45]*, the robot has phototaxis movement: it approaches, avoids, or remains stationary depending on the color of light. As described in *[35]*, the output *y* is velocity of the robot. There are five inputs: $u_1$ and $u_2$ are the absolute positions of the robot and the light, respectively, while $\{u_3, u_4, u_5\}$ is a binary, one-hot encoding of the light color, where 0 indicates the light is off and 1 indicates the light is on.

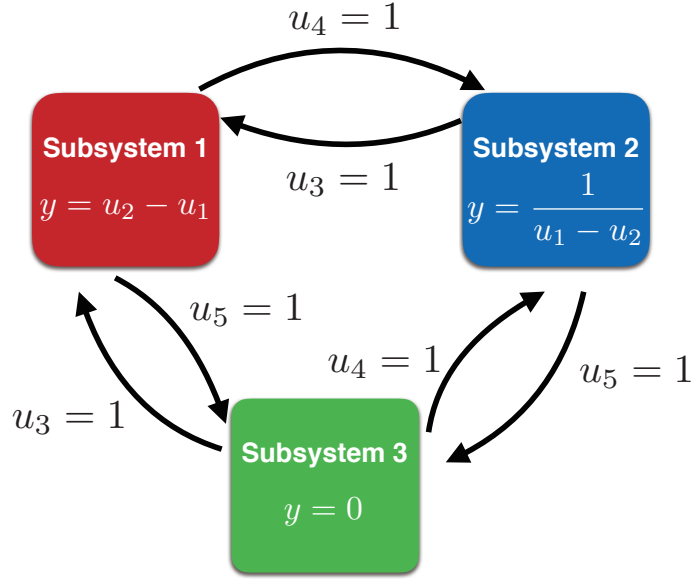

**Fig. S5.** The hybrid dynamical system model of Phototaxis Robot.

Similar to previous examples, 2000 data points are used. IHYDE starts obtaining false classification results as the noise level increases (shown in Table S9 and in Table S10). Yet, IHYDE is still able to identify the actual subsystem dynamics without redundant basis functions when noise intensity is low. When there exists redundant basis functions, IHYDE can identify all the subsystems when there is no noise. When noise level increases, IHYDE still identifies the right number of subsystems, yet the third identified subsystem is different from the true one, i.e., $y = 0$.

Again, once all subsystems have been identified and all data points have been classified, IHYDE identifies the transition logic between subsystems. IHYDE is able to precisely identify the correct transition logic both when there is no redundant basis function (Table S11) and when there is (Table S12). At a first glance, the inferred transition logic is different from the actual ones. Given $u_3, u_4, u_5$ are binary values, still, the inferred transition logic are equivalent to the actual ones.

**Table S9.** The identified result and details of tuning parameters in the Phototaxis Robot example.

| Noise | $N_p = 0$ | $N_p = 2\%$ | $N_p = 4\%$ | $N_p = 6\%$ |
|---|---|---|---|---|
| Library $\Phi$ | | $u_1 - u_2 \quad \frac{1}{u_1 - u_2}$ | | |
| Identified subsystem 1 | $y = -0.9980$ $(u_1 - u_2)$ | $y = -0.9978$ $(u_1 - u_2)$ | $y = -0.9964$ $(u_1 - u_2)$ | $y = -0.99555$ $(u_1 - u_2)$ |
| Identified subsystem 2 | $y = \frac{0.9908}{u_1 - u_2}$ | $y = \frac{0.9947}{u_1 - u_2}$ | $y = \frac{0.9820}{u_1 - u_2}$ | $y = \frac{0.9821}{u_1 - u_2}$ |
| Identified subsystem 3 | $y = 0$ | $y = 0$ | $y = 0.0068(u_1 - u_2)$ | $y = 0.0095(u_1 - u_2)$ |
| $\lambda_z$ | $5e-4$ | $5e-4$ | $5e-4$ | $0.001$ |
| $\varepsilon_z$ | | $1e-4$ | | |
| $\lambda_w$ | $0.05$ | $0.05$ | $0.1$ | $0.1$ |
| $\varepsilon_w$ | $0.05$ | $0.06$ | $0.2$ | $0.2$ |
| Number of misclassified points | $0$ | $11$ | $28$ | $47$ |

**Table S10.** The identified result and details of tuning parameters in the Phototaxis Robot example with redundant basis functions.

| Noise | $N_p = 0$ | $N_p = 2\%$ | $N_p = 4\%$ | $N_p = 6\%$ |
|---|---|---|---|---|
| Library $\Phi$ | | $1 \quad u_1 - u_2 \quad \frac{1}{u_1 - u_2} \quad u_1^2 \quad u_2^2$ | | |
| Identified subsystem 1 | $y = -1.0000$ $(u_1 - u_2)$ | $y = -0.9957$ $(u_1 - u_2)$ | $y = -0.9944$ $(u_1 - u_2)$ | $y = -0.9941$ $(u_1 - u_2)$ |
| Identified subsystem 2 | $y = \frac{0.9998}{u_1 - u_2}$ | $y = \frac{0.9891}{u_1 - u_2}$ | $y = \frac{0.9727}{u_1 - u_2}$ | $y = \frac{0.9689}{u_1 - u_2}$ |
| Identified subsystem 3 | $y = 0$ | $y = -0.0002 u_2^2$ | $y = 0.0046(u_1 - u_2)$ $+ 0.0014 u_1^2$ | $y = 0.0064(u_1 - u_2)$ $+ 0.0019 u_1^2$ |
| $\lambda_z$ | $1e-4$ | $1e-4$ | $5e-4$ | $1e-3$ |
| $\varepsilon_z$ | | $1e-4$ | | |
| $\lambda_w$ | $1e-3$ | $0.1$ | $0.15$ | $0.15$ |
| $\varepsilon_w$ | $0.005$ | $0.06$ | $0.2$ | $0.2$ |
| Number of misclassified points | $0$ | $11$ | $28$ | $48$ |

**Table S11.** The identified result and details of tuning parameters in the Phototaxis Robot example.

| System | Subsystem 1 | Subsystem 2 | Subsystem 3 |
|---|---|---|---|
| Subsystem 1 |  | $u_3 < 0.4986 u_4$ | $u_3 < 0.5085 u_5$ |
| Subsystem 2 | $u_4 < 0.4553 u_3$ |  | $u_4 < 0.5055 u_5$ |
| Subsystem 3 | $u_5 < 0.5242 u_3$ | $u_5 < 0.4543 u_4$ |  |
| Library $\Psi$ | 1 $u_3$ $u_4$ $u_5$ | | |
| $\beta$ | 0.5 | | |

**Table S12.** The identified transition logic for the Phototaxis Robot example.

| System | Subsystem 1 | Subsystem 2 | Subsystem 3 |
|---|---|---|---|
| Subsystem 1 |  | $u_3 < 0.4986 u_4$ | $u_3 < 0.5085 u_5$ |
| Subsystem 2 | $u_4 < 0.4553 u_3$ |  | $u_4 < 0.5055 u_5$ |
| Subsystem 3 | $u_5 < 0.5242 u_3$ | $u_5 < 0.4543 u_4$ |  |
| Library $\Psi$ | 1 $u_1^{-1}$ $u_2$ $\sin(u_1)$ $\cos(u_2)$ $e^{u_1 u_2}$ $u_3$ $u_4$ $u_5$ | | |
| $\beta$ | 0.5 | | |

## S3.4 Example 4: Non-linear Hybrid System

Consider the Nonlinear Hybrid System shown in Fig. S6. This example is a system without any physical counterpart, yet it is useful to evaluate the capabilities of IHYDE for finding non-linear expressions. The system consists of three subsystems, where all of the behaviors and transition logic consist of non-linear equations which cannot be modeled via parametric regression without incorporating prior knowledge. All the expressions are a function of the variables $u_1$ and $u_2$, the discriminant functions are not linearly separable and the transitions are modally dependent.

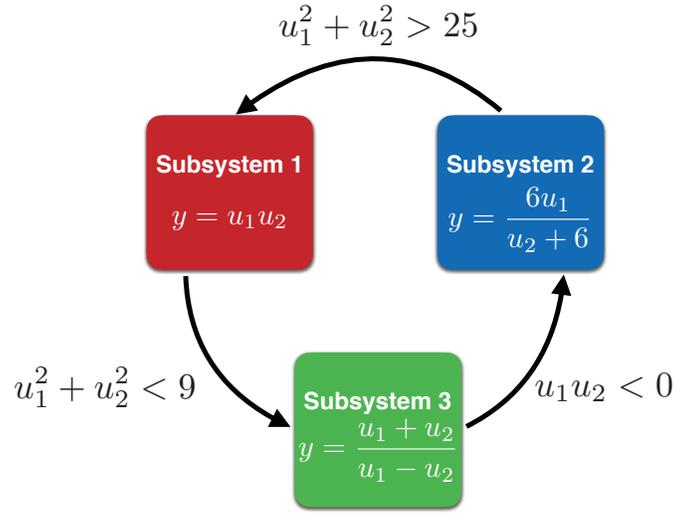

**Fig. S6.** The nonlinear hybrid dynamical system model.

Using 2000 data points in dataset generated by *[35]*, the identified results are shown in Table S13 and Table S14. Using IHYDE for subsystems identification, we are able to successfully identify that there are three subsystems that generate the datasets. In addition, the three identified subsystems are consistent with or close to the true ones from both noiseless and noisy data. IHYDE is also able to precisely identify the correct transition logic both when there is no redundant basis function (Table S15) and when there is (Table S16).

**Table S13.** The identified result and details of Non-linear Hybrid System.

| Noise | $N_p = 0$ | $N_p = 2\%$ | $N_p = 4\%$ | $N_p = 6\%$ |
|---|---|---|---|---|
| Library $\Phi$ | | $u_1 u_2 \quad \frac{6u_1}{6+u_2} \quad \frac{u_1+u_2}{u_1-u_2}$ | | |
| Identified subsystem 1 | $y = 0.9998 u_1 u_2$ | $y = 0.9958 u_1 u_2$ | $y = 0.9962 u_1 u_2$ | $y = 0.9953 u_1 u_2$ |
| Identified subsystem 2 | $y = 0.9983 \frac{6u_1}{6+u_2}$ | $y = 0.9961 \frac{6u_1}{6+u_2}$ | $y = 0.9947 \frac{6u_1}{6+u_2}$ | $y = 0.9939 \frac{6u_1}{6+u_2}$ |
| Identified subsystem 3 | $y = 0.9990 \frac{u_1+u_2}{u_1-u_2}$ | $y = 0.9945 \frac{u_1+u_2}{u_1-u_2}$ | $y = 0.9940 \frac{u_1+u_2}{u_1-u_2}$ | $y = 0.9949 \frac{u_1+u_2}{u_1-u_2}$ |
| $\lambda_z$ | $1e-6$ | $1.5e-4$ | $1.5e-4$ | $1.5e-4$ |
| $\varepsilon_z$ | $1e-4$ | | | |
| $\lambda_w$ | 0.03 | | | |
| $\varepsilon_w$ | 0.6 | 2 | 2 | 2 |
| Number of misclassified points | 0 | 63 | 129 | 177 |

**Table S14.** The identified result and details of Non-linear Hybrid Systems when there exists redundant basis functions.

| Noise | $eps = 0$ | $eps = 2\%$ | $eps = 4\%$ | $eps = 6\%$ |
|---|---|---|---|---|
| Library $\Phi$ | $u_1 u_2 \quad \frac{6u_1}{6+u_2} \quad \frac{u_1+u_2}{u_1-u_2} \quad u_1 \quad u_2 \quad \sin(u_1) \quad \sin(u_2) \quad u_1^2 \quad u_2^2$ | | | |
| Identified subsystem 1 | $y = 0.9997 u_1 u_2$ | $y = 0.9999 u_1 u_2$ | $y = 0.9944 u_1 u_2$ | $y = 0.9953 u_1 u_2$ |
| Identified subsystem 2 | $y = 0.9983 \frac{6u_1}{6+u_2}$ | $y = 0.9960 \frac{6u_1}{6+u_2}$ | $y = 0.9960 \frac{6u_1}{6+u_2}$ | $y = 0.9963 \frac{6u_1}{6+u_2}$ |
| Identified subsystem 3 | $y = 0.9990 \frac{u_1+u_2}{u_1-u_2}$ | $y = 0.9975 \frac{u_1+u_2}{u_1-u_2}$ | $y = 0.9898 \frac{u_1+u_2}{u_1-u_2}$ | $y = 0.9877 \frac{u_1+u_2}{u_1-u_2}$ |
| $\lambda_z$ | $1e-5$ | $5e-5$ | $1.5e-4$ | $1.5e-4$ |
| $\varepsilon_z$ | $1e-4$ | | | |
| $\lambda_w$ | 0.03 | 0.03 | 0.032 | 0.0282 |
| $\varepsilon_w$ | 0.6 | 0.8 | 2 | 2 |
| Number of misclassified points | 0 | 67 | 129 | 175 |

**Table S15.** The identified transition logic of Non-linear Hybrid System.

| System | Subsystem 1 | Subsystem 2 | Subsystem 3 |
|---|---|---|---|
| Subsystem 1 | | | $u_1^2 + u_2^2 < 8.9993$ |
| Subsystem 2 | $u_1^2 + u_2^2 > 24.9803$ | | |
| Subsystem 3 | | $u_1 u_2 < -0.013$ | |
| Library $\Psi$ | $1 \quad u_1 u_2 \quad u_1^2 + u_2^2$ | | |
| $\beta$ | 0.01 | | |

**Table S16.** The identified transition logic of Non-linear Hybrid System when there are redundant basis functions.

| System | Subsystem 1 | Subsystem 2 | Subsystem 3 |
|---|---|---|---|
| Subsystem 1 | | | $30.8698u_1^2 + 31.9412u_2^2$ $< 282.0993$ |
| Subsystem 2 | $11.7323u_1^2 + 11.7384u_2^2$ $> 292.7959$ | | |
| Subsystem 3 | | $u_1 u_2 < -0.013$ | |
| Library $\Psi$ | \multicolumn{3}{c}{$1 \quad u_1 \quad u_2 \quad e^{u_1+u_2} \quad u_1 u_2 \quad u_1^2 \quad u_2^2$} |
| $\beta$ | \multicolumn{3}{c}{0.01} |

## S3.5 Example 5: Power Grid Fault Detection

The next example illustrates how IHYDE can be used in real-time monitoring applications. Consider the fault detection problem in a smart grid. The design of monitoring schemes to diagnose anomalies caused by unpredicted or sudden faults on power networks is of great importance.

Here we consider a benchmark power network in Fig. S7. Suppose the line connecting buses 6 and 12 disconnects at time 31, changing the admittance between these two buses to zero. We simulate the data and only pass the data to IHYDE without other information. IHYDE can immediately detect the occurrence of this event and estimate the new admittance matrix using measurements of the next 10 measurements. It successfully discovers two different subsystems from data and pinpoints the difference in the discovered subsystems which corresponds to the fault. Given the frequency at which PMUs sample voltage and current, IHYDE is able to locate the fault in a few hundred milliseconds after the event occurs, enabling the operators to detect the event, identify its location, and take remedial actions in near real-time.

**Table S17.** The identified result and details of power system fault detection.

| Bus | Bus 6 and bus 12 | Other bus except bus 1 | Bus 1 |
|---|---|---|---|
| True time for fault occurance | 31 | None | None |
| Identified time for fault occurance | 31 | None | None |
| $\lambda_z$ | $1e-3$ | $1e-3$ | $1e-3$ |
| $\varepsilon_z$ | 0.008 | 0.008 | 0.008 |
| $\lambda_w$ | $1e-6$ | $1e-6$ | $1e-9$ |
| $\varepsilon_w$ | 0.05 | 0.05 | 0.05 |
| Library $\Psi$ | $1\ t$ | | |
| $\beta$ | 0.01 | None | None |

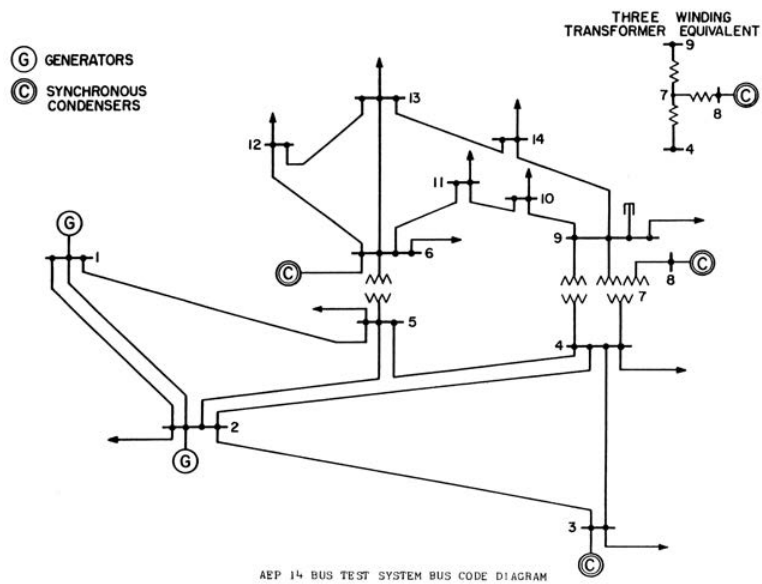

**Fig. S7.** The grid topology (IEEE 14-bus test case) that we used to simulate the data. Figure is obtained from http://www2.ee.washington.edu/research/pstca/pf14/pg_tca14bus.html. We used MATPOWER to simulate the system to obtain the data. All codes are available.

## S3.6  Example 6: Identification of Real-time Models for Smart Grid

This example illustrates how the proposed IHYDE method can be used to solve the identification problem in smart grid, which contains two major parts, smart infrastructure system and smart management system *[46]*. It is crucial to obtain real-time models for smart management system to achieve resilient and efficient operations. Accurate model information is not only necessary for daily operation and scheduling, but also critical for other advanced techniques such as state estimation and optimal power flow computation. However, such information is not always available in distribution systems due to frequent model changes. For example, the model of a distribution system connected with photovoltaic panels maybe change once every eight hours *[47]*. Furthermore, some unexpected events, such as line faults and unreported line maintenance, can lead to model changes. Moreover, network reconfiguration (such as switch action for balancing loads and avoiding voltage sag) happens frequently in distribution systems. Therefore, model identification in real-time is meaningful.

We apply IHYDE to identify network models in real-time and to infer transition logic for model changes using data from advanced metering infrastructure. The 33-bus benchmark distribution system *[32]* shown in Figure S8 is used to generate data. Consider the situation where loads at some remote nodes of a feeder suddenly increase causing voltage sag, subsequently, an operator takes switch action for load balancing and voltage regulation. Figure S9 depicts the switching topologies and the real transition logic. The detailed actions and switching time are shown in Table S18. Measurements are generated via solving nonlinear power flow equations using MATPOWER toolbox *[48]* in MATLAB.

Suppose that we can measure all the active and reactive power consumption, and voltage phasors of the nodes, denoted by *M* as follows.

$$M = \begin{bmatrix} P_1^{(1)} & Q_1^{(1)} & U_1^{(1)} & \cdots & P_{33}^{(1)} & Q_{33}^{(1)} & U_{33}^{(1)} \\ \vdots & \vdots & \vdots & \vdots & \vdots & \vdots & \vdots \\ P_1^{(m)} & Q_1^{(m)} & U_1^{(m)} & \cdots & P_{33}^{(m)} & Q_{33}^{(m)} & U_{33}^{(m)} \end{bmatrix} \quad (10)$$

where $U_i^{(m)}$, $P_i^{(m)}$ and $Q_i^{(m)}$ are the voltage phasor, active and reactive power of Bus *i* at time instant *m*, respectively. The total sampling time *m* is set to 180 in the following simulation and measurement noise is not considered.

For each node, we apply IHYDE to identify the responding column of the admittance matrix. The output $y_i \in R^{2m \times 1}$ of Bus *i* is constructed according to its active and reactive power as $y_i =$

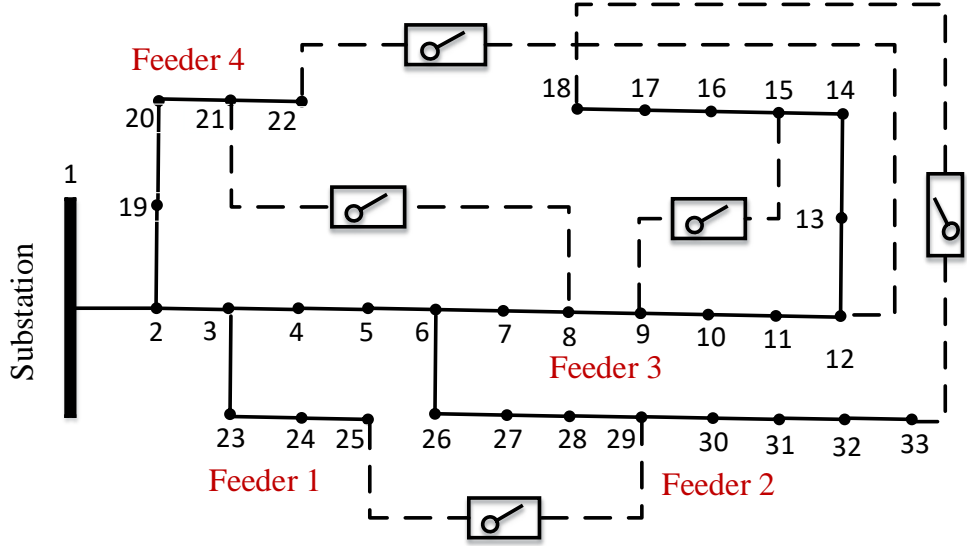

**Fig. S8.** Diagram of the 33-bus distribution system with four feeders. Every node represents a concentrated load with active and reactive power consumption. The solid line stands for the electric line connecting two distinct nodes in service with close sectionalizing switch, and dotted lines represent the open tie switches.

$[P_i^{(1)}, Q_i^{(1)}, \cdots, P_i^{(m)}, Q_i^{(m)}]^T$. Its dictionary matrix $\Phi_i \in R^{2m \times 66}$ can be formulated as follows.

$$\Phi_i = \begin{bmatrix} |U_i^{(1)}||U_1^{(1)}|\cos\theta_{i1}^{(1)} & \cdots & |U_i^{(1)}||U_{33}^{(1)}|\cos\theta_{i33}^{(1)} & |U_i^{(1)}||U_1^{(1)}|\sin\theta_{i1}^{(1)} & \cdots & |U_i^{(1)}||U_{33}^{(1)}|\sin\theta_{i33}^{(1)} \\ |U_i^{(1)}||U_1^{(1)}|\sin\theta_{i1}^{(1)} & \cdots & |U_i^{(1)}||U_{33}^{(1)}|\sin\theta_{i33}^{(1)} & -|U_i^{(1)}||U_1^{(1)}|\cos\theta_{i1}^{(1)} & \cdots & -|U_i^{(1)}||U_{33}^{(1)}|\cos\theta_{i33}^{(1)} \\ \vdots & \vdots & \vdots & \vdots & \vdots & \vdots \\ |U_i^{(m)}||U_1^{(m)}|\cos\theta_{i1}^{(m)} & \cdots & |U_i^{(m)}||U_{33}^{(m)}|\cos\theta_{i33}^{(m)} & |U_i^{(m)}||U_1^{(m)}|\sin\theta_{i1}^{(m)} & \cdots & |U_i^{(m)}||U_{33}^{(m)}|\sin\theta_{i33}^{(m)} \\ |U_i^{(m)}||U_1^{(m)}|\sin\theta_{i1}^{(m)} & \cdots & |U_i^{(m)}||U_{33}^{(m)}|\sin\theta_{i33}^{(m)} & -|U_i^{(m)}||U_1^{(m)}|\cos\theta_{i1}^{(m)} & \cdots & -|U_i^{(m)}||U_{33}^{(m)}|\cos\theta_{i33}^{(m)} \end{bmatrix}$$

where $|U_i^{(m)}|$ and $\theta_{ij}^{(m)}$ denote the voltage magnitude of Bus $i$ and the phase difference between nodal voltages of Bus $i$ and $j$ at time instant $m$, respectively.

Table S19 shows the identified results and the detailed tuning parameters of the proposed al-

**Table S18.** The detailed parameters of switch operators.

| Model Switching | Time | Opened Switch | Closed Switch | Bus of load change |
|---|---|---|---|---|
| $\mathscr{T}_1 \to \mathscr{T}_2$ | 31, 91, 151 | $11-12$ | $12-22$ | 9, 10, 11 |
| $\mathscr{T}_2 \to \mathscr{T}_1$ | 61, 121 | $12-22$ | $11-12$ | 20, 21, 22 |

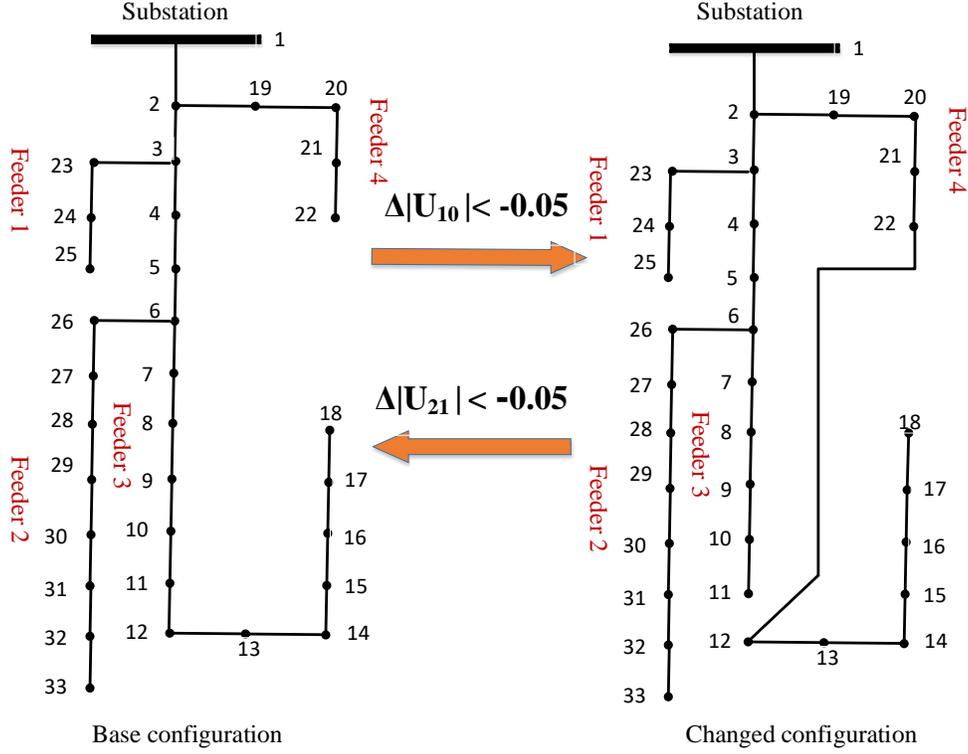

**Fig. S9.** Subsystem models and transition logic of the smart grid example.

gorithm. For example, at Bus 12, the maximum relative identification ratio of Base configuration and Changed configuration are 0.00057% and 0.00182%, respectively. The identified admittance matrices at time instants $31, 61, 91, 121, 151$ are very different from that of the previous moments, which indicates the model switching. The results demonstrate that IHYDE can identify the models accurately and pinpoint model switching time correctly. We add the difference of voltage magnitude between different times, denoted by $\Delta|U_i|$, into dictionary matrix for logic identification. Table S20 indicates that the identified logic is consistent with the real logic with small error. Specifically, the result of $\mathcal{T}_1 \to \mathcal{T}_2$ (switching from $\mathcal{T}_1$ to $\mathcal{T}_2$) reveals that the voltage drop of node 10 at feeder 3 are more than 0.0500 at time 30, subsequently, switch action is taken to avoid sharp voltage drop. The tie switch between Bus 12 and 22 is closed, while the sectionalizing switch between Bus 11 and 12 opens. This is consistent with our preset reason that loads at Bus $9, 10, 11$ increase rapidly at time 30. There are many indistinct physical phenomenons in actual power system and IHYDE can be utilized to help engineers understand the hidden mechanism behind it.

**Table S19.** The identified result and detailed parameters.

| System model | $\mathscr{T}_1 \to \mathscr{T}_2$ | $\mathscr{T}_2 \to \mathscr{T}_1$ |
|---|---|---|
| True switching time | 31, 91, 151 | 61, 121 |
| Identified switching time | 31, 91, 151 | 61, 121 |
| $\lambda_z$ | \multicolumn{2}{c}{$5e-3$} |
| $\varepsilon_z$ | \multicolumn{2}{c}{$1.5e-2$} |
| $\lambda_w$ | \multicolumn{2}{c}{$1e-6$} |
| $\varepsilon_w$ | \multicolumn{2}{c}{$5e-2$} |
| Number of misclassified points | \multicolumn{2}{c}{0} |

**Table S20.** The identified transition logic for the model switching in smart grid.

| System | Subsystem 1 | Subsystem 2 |
|---|---|---|
| Subsystem 1 | | $\Delta|U_{10}| < -0.0500$ |
| Subsystem 2 | $\Delta|U_{21}| < -0.0473$ | |
| Library $\Psi$ | \multicolumn{2}{c}{$1\ \Delta|U_1|\cdots\Delta|U_{33}|$} |
| $\beta$ | \multicolumn{2}{c}{$1e-4$} |

## S3.7 Example 7: Discovery of Human Atrial Action Potential Models

In this section, we apply IHYDE to a human atrial action potential (AP) model proposed in *[34]* to show the applicability of IHYDE to the discovery in biology. The parameters of the human atrial AP model are determined based on the data that is directly measured on human atrial cells and that is from AP model of guinea pig ventricular and rabbit atrial. The AP model can reproduce a variety of observed AP behaviors and provide potential insights into its underlying ionic mechanisms. The human atrial AP and ionic currents that underlie its morphology are of great importance to our understanding and prediction of the electrical properties of atrial tissues under normal and pathological conditions.

Specifically, the cell membrane is modeled as a capacitor connected in parallel with variable resistances and batteries representing the ionic channels and driving forces. The AP model includes 21 differential equations and 163 parameters in total (see the code for detailed information). The membrane potential formulation is as follows

$$\frac{dV}{dt} = \frac{-(I_{ion} + I_{st})}{C_m}, \tag{11}$$

where $V$ is membrane potential, and $C_m$ is the constant total membrane capacitance. $I_{ion}$ and $I_{st}$ are the total ionic current and stimulus current flowing across the membrane, respectively.

Figure S10 shows that the action potential generated by the AP model through voltage clamp method is a spike-and-dome morphology commonly observed in human atrial AP recordings. We apply the stimulation current with 2 ms pulses of 2 nA amplitude across the cell membrane every 1000 ms. To check the performance of the IHYDE method, we focus on two representative equations about gating variables $h$ and $j$ with time-varying parameters as follows:

$$\frac{dh}{dt} = \alpha_h - h(\alpha_h + \beta_h) \tag{12}$$

$$\frac{dj}{dt} = \alpha_j - j(\alpha_j + \beta_j), \tag{13}$$

where $h$ and $j$ are fast and slow inactivation gating variables for fast inward $Na^+$ current, respectively. For convenience, we present the time-varying parameters $\alpha_h, \alpha_j, \beta_h, \beta_j$:

$$\alpha_h = \begin{cases} \alpha_{h1} \triangleq 0.135 \exp(-\frac{V+80}{6.8}) & V < -40 \\ \alpha_{h2} \triangleq 0 & V \geq -40 \end{cases}$$

49

$$\alpha_j = \begin{cases} \alpha_{j1} \triangleq [-1.2714 \times 10^5 \exp(0.2444V) - \\ \qquad 3.474 \times 10^{-5} \exp(-0.04391)] \frac{V+37.78}{1+\exp[0.311(V+79.23)]} & V < -40 \\ \alpha_{j2} \triangleq 0 & V \geq -40 \end{cases}$$

$$\beta_h = \begin{cases} \beta_{h1} \triangleq 3.56\exp(0.079V) + 3.1 \times 10^5 \exp(0.35V) & V < -40 \\ \beta_{h2} \triangleq \{0.13[1+\exp(-\frac{V+10.66}{11.1})]\}^{-1} & V \geq -40 \end{cases}$$

$$\beta_j = \begin{cases} \beta_{j1} = 0.1212 \frac{\exp(-0.01052V)}{1+\exp[-0.1378(V+40.14)]} & V < -40 \\ \beta_{j2} = 0.3 \frac{\exp(-2.535 \times 10^{-7}V)}{1+\exp[-0.1(V+32)]} & V \geq -40. \end{cases}$$

When the gating variables $h$ and $j$ are equal to 1, the fast inward $Na^{2+}$ current is inactive completely. Figure S11 depicts that they gradually rise to their resting values 0.9775 and 0.9649 after stimulus.

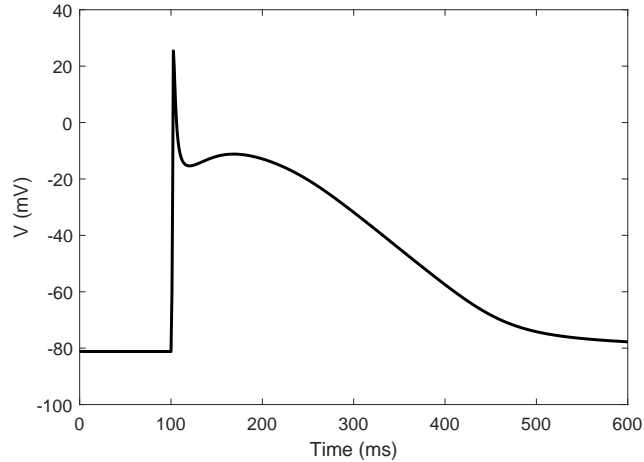

**Fig. S10.** Model action potential $V$ during stimulation at the frequency of 1 Hz.

It is clearly observed that membrane voltage gradually returns to its stable resting potential $-81$mV after the stimulation from Figure S10. During the process, the dynamics for gating variables $h$ and $j$ has been switched as shown in Figure S11 when the membrane voltage $V$ goes through $-40$ mV. We apply IHYDE to discover the different models and the transition logic only using measurements. The first-order differential values of $h$ and $j$ are considered as their output, respectively. For instance, we down-sample the differential value of $h$ during $120-500$ ms as its output

$$y_h = \left[\frac{dh(120)}{dt}, \frac{dh(120+\delta_t)}{dt}, \cdots, \frac{dh(499.8)}{dt}\right]^T \in R^{1267 \times 1}.$$

The sampling period $\delta_t$ is set to 0.3 ms, and there are 1267 data points for each variable. The

50

dictionary matrix of gating variables $h$ and $j$, denoted by $\Phi_h$ and $\Phi_j$, respectively, are established based on the terms of the above equations

$$\Phi_h = \begin{bmatrix} \exp(-\frac{V(t_0)+80}{6.8}) & h(t_0)\exp(-\frac{V(t_0)+80}{6.8}) & h(t_0)\beta_{h1}(t_0) & h(t_0)\beta_{h2}(t_0) \\ \vdots & \vdots & \vdots & \vdots \\ \exp(-\frac{V(t_1)+80}{6.8}) & h(t_1)\exp(-\frac{V(t_1)+80}{6.8}) & h(t_1)\beta_{h1}(t_1) & h(t_1)\beta_{h2}(t_1) \end{bmatrix},$$

$$\Phi_j = \begin{bmatrix} \alpha_{j1}(t_0) & j(t_0)\alpha_{j1}(t_0) & j(t_0)\frac{\exp(-0.01052V(t_0))}{1+\exp[-0.1378(V(t_0)+40.14)]} & j(t_0)\frac{\exp(-2.535\times 10^{-7}V(t_0))}{1+\exp[-0.1(V(t_0)+32)]} \\ \vdots & \vdots & \vdots & \vdots \\ \alpha_{j1}(t_1) & j(t_1)\alpha_{j1}(t_1) & j(t_1)\frac{\exp(-0.01052V(t_1))}{1+\exp[-0.1378(V(t_1)+40.14)]} & j(t_1)\frac{\exp(-2.535\times 10^{-7}V(t_1))}{1+\exp[-0.1(V(t_1)+32)]} \end{bmatrix},$$

where $t_0$ and $t_1$ are 120 and 499.8 ms, respectively.

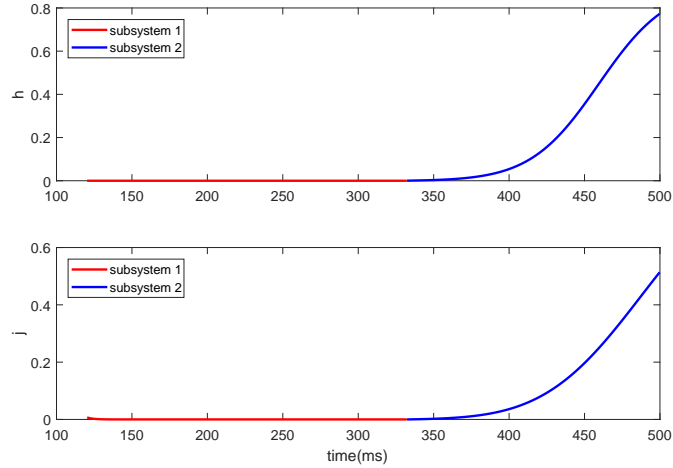

**Fig. S11.** The value of gating variable $h$ and $j$. Different colors denote data that are produced from different subsystems.

The identified results and the detailed parameters are summarized in Table S21. We can see that IHYDE identifies the subsystem and pinpoints the changing time correctly. The identified logic for both gating variables are $V \leq -40.0093$, which is very close to the real logic $V \leq -40$.

Table S21. The identified results and detailed parameters.

| Gating variable | $h$ | $j$ |
|---|---|---|
| Actual subsystem 1 | $\dot{h} = -h\beta_{h2}$ | $\dot{j} = -0.3j\frac{\exp(-2.535\times 10^{-7}V)}{1+\exp[-0.1(V+32)]}$ |
| Actual subsystem 2 | $\dot{h} = 0.135\exp(-\frac{V+80}{6.8}) - 0.135h\exp(-\frac{V+80}{6.8}) - h\beta_{h1}$ | $\dot{j} = \alpha_{j1} - j\alpha_{j1} - 0.1212j\frac{\exp(-0.01052V)}{1+\exp[-0.1378(V+40.14)]}$ |
| Actual change time | 332.10 ms | 332.10 ms |
| Identified subsystem 1 | $\dot{h} = -0.9999h\beta_{h2}$ | $\dot{j} = -0.3000j\frac{\exp(-2.535\times 10^{-7}V)}{1+\exp[-0.1(V+32)]}$ |
| Identified subsystem 2 | $\dot{h} = 0.1349\exp(-\frac{V+80}{6.8}) - 0.1349\exp(-\frac{V+80}{6.8})h - 0.9987h\beta_{h1}$ | $\dot{j} = 1.0000\alpha_{j1} - 1.0000j\alpha_{j1} - 0.1212j\frac{\exp(-0.01052V)}{1+\exp[-0.1378(V+40.14)]}$ |
| Identified change time | 332.10 ms | 332.10 ms |
| $\lambda_z$ | 1e-4 | 1e-4 |
| $\varepsilon_z$ | 3e-5 | 3e-5 |
| $\lambda_w$ | 3e-5 | 1e-5 |
| $\varepsilon_w$ | 5e-5 | 5e-5 |
| Library $\Phi$ | $\exp(-\frac{V+80}{6.8})$  $h\exp(-\frac{V+80}{6.8})$  $h\beta_{h1}$  $h\beta_{h2}$ | $\alpha_{j1}$  $j\alpha_{j1}$  $j\frac{\exp(-0.01052V)}{1+\exp[-0.1378(V+40.14)]}$  $j\frac{\exp(-2.535\times 10^{-7}V)}{1+\exp[-0.1(V+32)]}$ |
| Number of misclassfied points | 0 | 0 |

Table S22. The identified transition logic for gating variable $h$.

| gating variable $h$ | Subsystem 1 | Subsystem 2 |
|---|---|---|
| Subsystem 1 |  | $V \leq -40.0093$ |
| Library $\Psi$ |  | 1  $V$ |
| $\beta$ |  | 1e-6 |

Table S23. The identified transition logic for gating variable $j$.

| gating variable $j$ | Subsystem 1 | Subsystem 2 |
|---|---|---|
| Subsystem 1 | | $V \leq -40.0093$ |
| Library $\Psi$ | 1 $V$ | |
| $\beta$ | 1e-6 | |

## S3.8 Example 8: Monitoring of Industrial Processes

The next example illustrates how IHYDE can be used for fault detection in mechanical engineering. Experiments conducted on a wind turbine system platform *[30]* shown in Figure S12 and S13 are used to verify its effectiveness.

This system contains a power supply of 380$V$, an inverter, a motor, a gearbox, a power generator, and a load. The platform is used to simulate the process of air flow through wind turbines to generate electricity. Specifically, the motor with a gear reducer of 20 : 1 ratio supplies the generator with mechanical power through the gearbox. In this experiment, we adopt the mode of gearbox inversion to simulate the operation of wind turbine system. The gearbox has been widely used to provide speed and torque conversions from a motor to generator in wind turbines *[49]*. This system has a gearbox with three shafts, i.e., shaft with low speed, shaft with intermediate speed and shaft with high speed. The load consumes the power generated by the generator. We can measure the root-mean-square current and voltage of the motor from the inverter. The current of generator can be captured by oscilloscope and its voltage is measured through multimeter. We measure the voltage of the load in the same way.

We perform experiments under normal and faulty conditions. Both experiments are performed in the situation where the generator speed is 200 revolutions per minute and the load is 1.5 KNm. One-third of the tooth width cut off from the gear tooth on the high-speed shaft is considered as the faulty condition shown in Figure S14. In the normal operation, the motor power is 383.01$W$ and the generator power is 53.28$W$; the load voltage is 75$V$ in the faulty condition.

Two sets of current data are measured at the frequency of 1000Hz connected in series for identification. The first dataset contains 19,995 data points sampled under normal operating condition, the other has 20,000 data points obtained from the faulty condition. Then, we down-sample at the period of 0.3s and denote as $i(k)$ in which $k = 1, \ldots, 133$.

As described in the main text, here we used an online monitoring scheme. We construct the output $y \in R^{64 \times 1}$, including 61 current measurements from 1.8s to 19.8s in the normal condition and 3 data points from 20.1s to 20.7s in the faulty condition when the mismatch is large. Specifically

$$y = \begin{bmatrix} i(7) & i(8) & \ldots & i(70) \end{bmatrix}^T.$$

The kth row of the dictionary matrix $\Phi \in R^{64 \times 28}$ is all the polynomial combinations of $i(k), \ldots, i(k+$

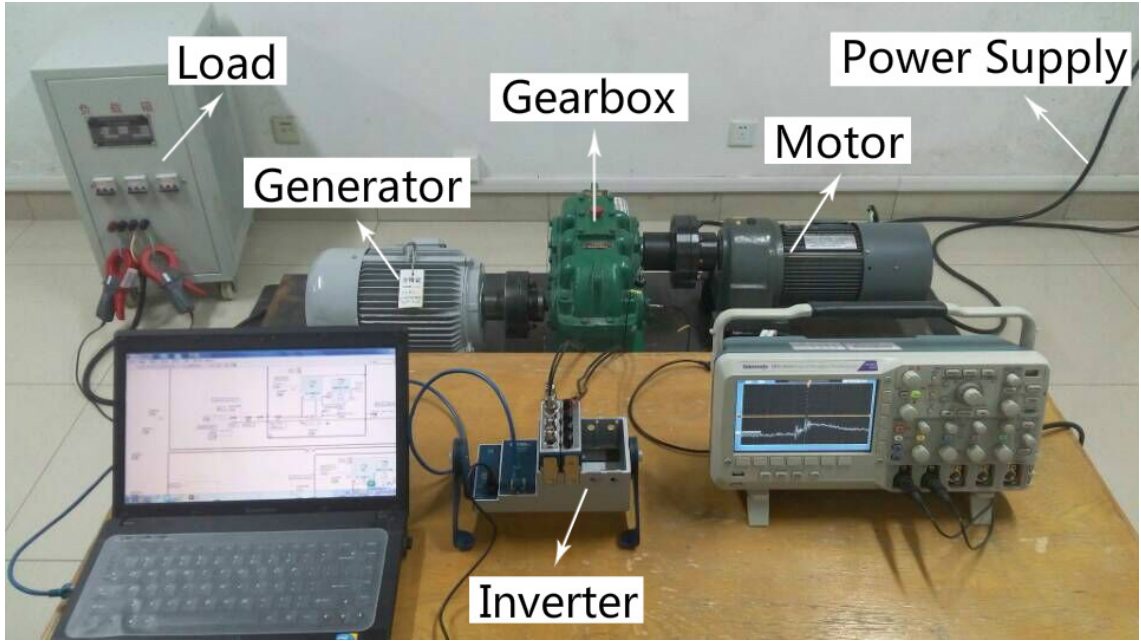

**Fig. S12.** A picture of the wind turbine system platform built in *[30]*.

5) to second order as follows:

$$\Phi = \begin{bmatrix} 1 & i(6) & \cdots & i(1) & i^2(6) & \ldots & i^2(1) \\ \vdots & \vdots & \vdots & \vdots & \vdots & \vdots & \vdots \\ 1 & i(69) & \ldots & i(64) & i^2(69) & \ldots & i^2(64) \end{bmatrix},$$

Figure S15 shows that the relative fitting error ratio is small. The identified results and details of the IHYDE are presented in Table S24, which shows that the identified time for the fault occurrence is the same as the real fault time 68. We only use three fault points to realize the fault detection. This example demonstrates the capability of IHYDE to the identification of the fault in industrial processes.

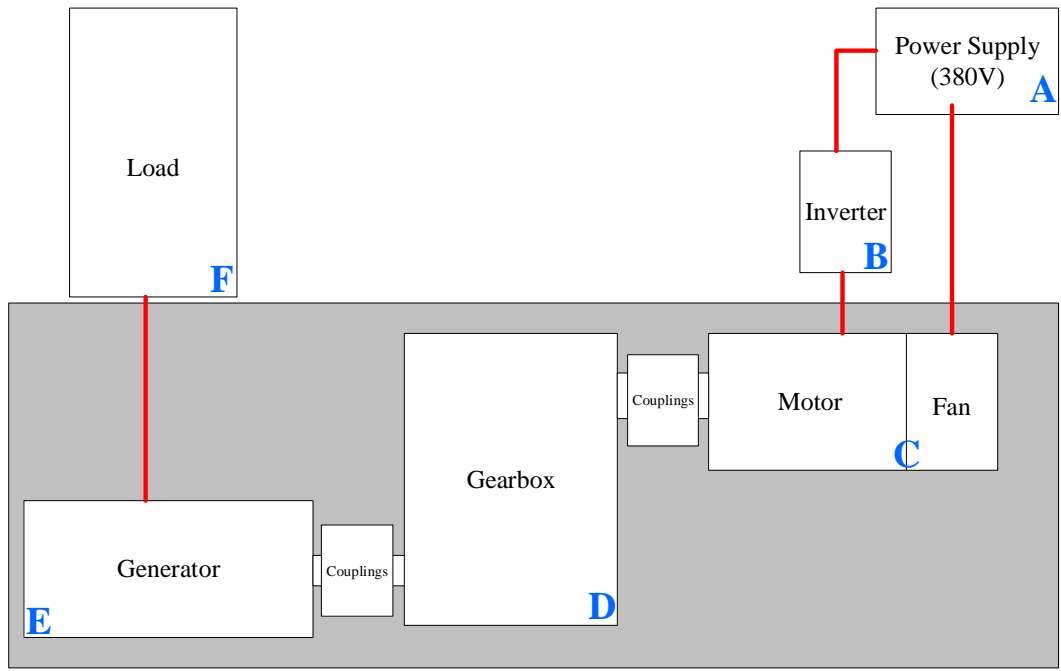

**Fig. S13.** The corresponding schematic diagram of the wind turbine system platform.

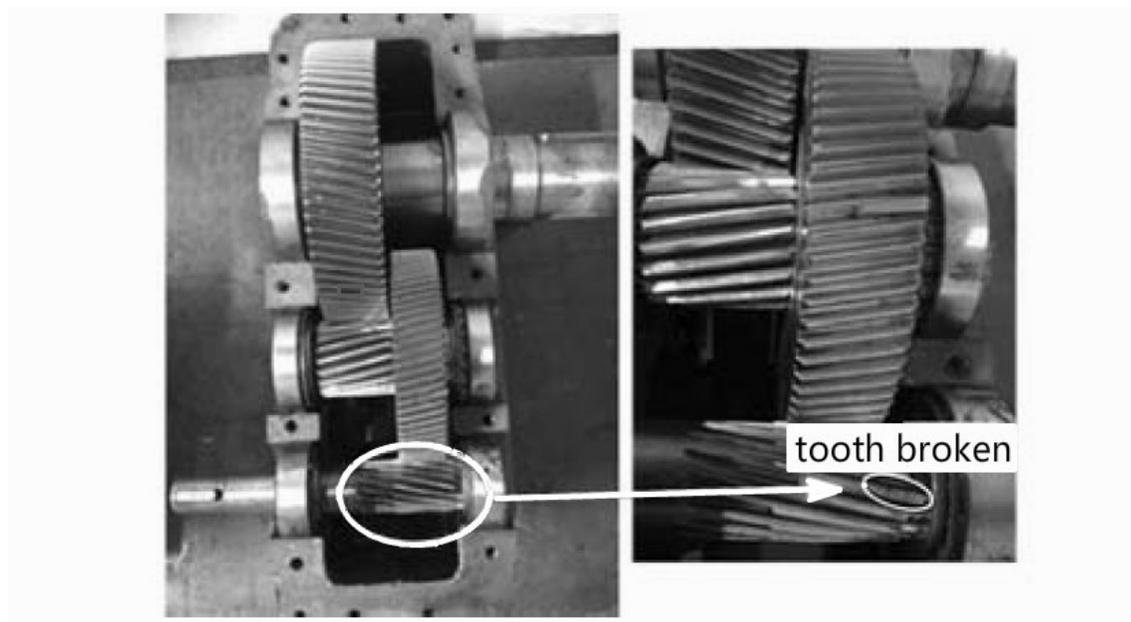

**Fig. S14.** The broken tooth in the gearbox *[30]*.

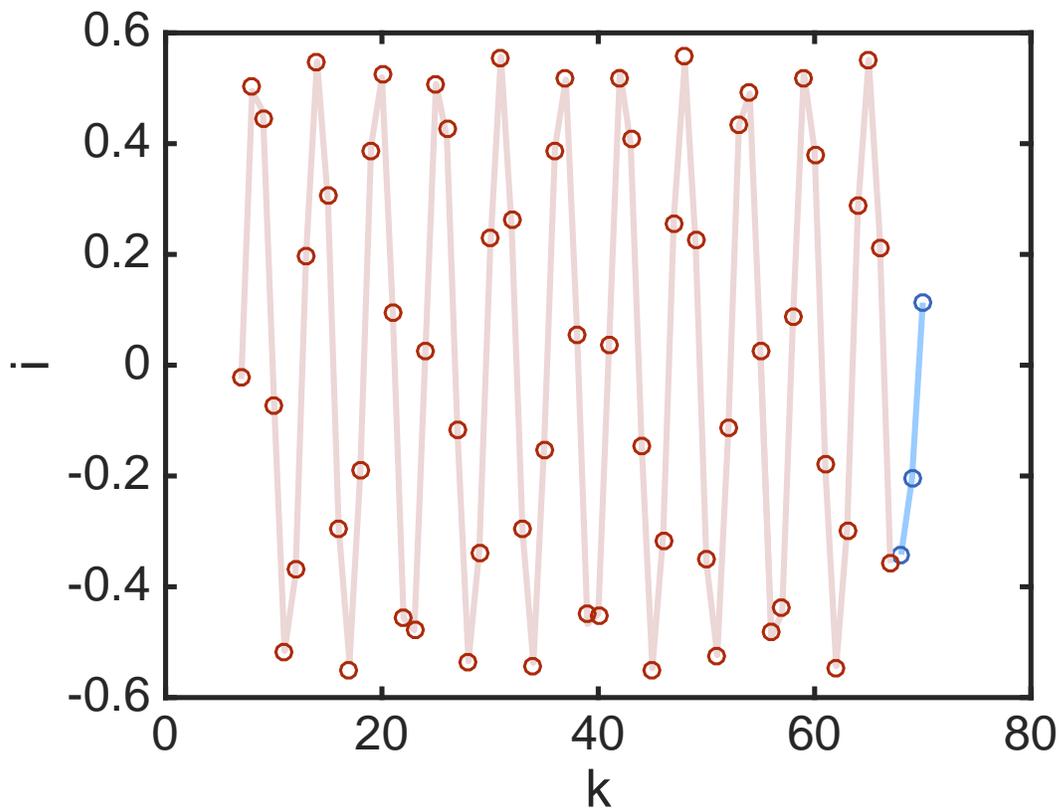

**Fig. S15.** The data fitting curve using data obtained from the wind turbine platform using IHYDE.The original time-series data (lines connecting the dots) in the color associated with the subsystem, the fitted data from the identified models (dots), and the location of the transitions (changes in colors).

Table S24. The identified result and details of the gearbox broken tooth fault with redundant basis functions.

| System | Gearbox |
|---|---|
| True fault time | 68 |
| Identified fault time | 68 |
| $\lambda_z$ | 1.5 |
| $\varepsilon_z$ | $1e-4$ |
| $\lambda_w$ | $5e-5$ |
| $\varepsilon_w$ | 0.026 |
| Library $\Phi$ | all the polynomial combinations of $y(k)\cdots y(k-5)$ to second order |
| Number of misclassified points | 0 |
| Library $\Psi$ | 1 $k$ |
| $\beta$ | 0.1 |

## S3.9 Example 9: Chua's Circuit

In this subsection and the next one, we shall apply IHYDE to data that is obtained from experiments. Chua's circuit shown in Fig. S16 is the simplest electronic circuit that exhibits classic chaotic behavior. It consists of an inductor, two capacitors, a passive resistor and an active nonlinear resistor which fits the condition for chaos with the least components. The most important active nonlinear resistor is a conceptual component and the resistor can be built with operational amplifiers and linear resistors. The current-voltage characteristics of the nonlinear resistor are plotted in Fig. S17.

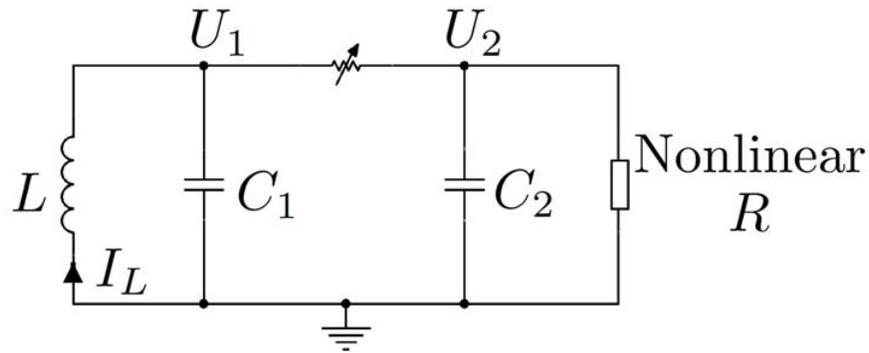

**Fig. S16.** The circuit structure of the Chua's circuit.

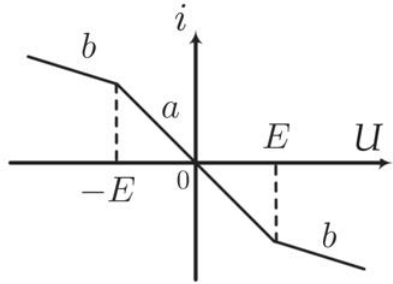

**Fig. S17.** The current-voltage characteristics of the nonlinear resistor.

By design, the current-voltage relationship can be described as follows:

$$i(U) = \begin{cases} aU + (b-a)(U-E), & U > E \\ aU, & -E < U < E \\ aU + (b-a)(U+E), & U < -E, \end{cases} \quad (14)$$

59

or equivalently

$$i(U) = bU + \frac{1}{2}(a-b)(|U+E| - |U-E|). \tag{15}$$

In both equations, $a$, $b$, $E$ are parameters defined in Fig. S17.

The nonlinear resistor can be built using the circuit realization as shown in Fig. S18.

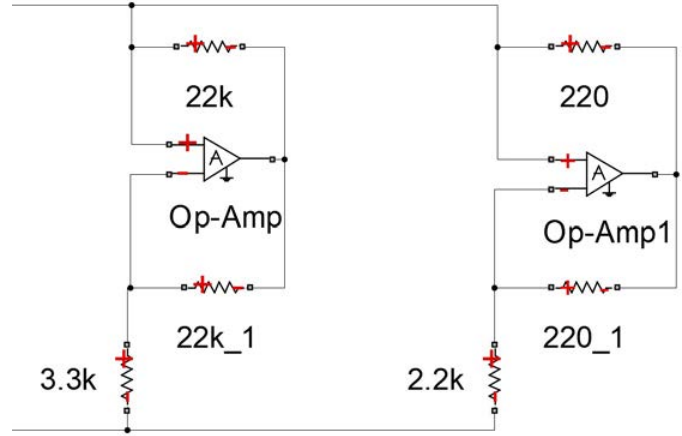

**Fig. S18.** The circuit structure of nonlinear resistor with specified current-voltage realization.

From KCL and KVL, we obtain

$$\begin{aligned}
\frac{C_1 dU_1}{dt} &= \frac{U_2 - U_1}{R} - i(U_1) \\
\frac{C_2 dU_2}{dt} &= \frac{U_1 - U_2}{R} + I_L \\
-L\frac{dI_L}{dt} &= U_2.
\end{aligned} \tag{16}$$

Then we introduce a number of variables to simplify the above equations:

$$\begin{aligned}
y_1 &= \frac{U_1 * \text{threshold}}{E} \\
y_2 &= \frac{U_2 * \text{threshold}}{E} \\
y_3 &= \frac{I_L * \text{threshold}}{E},
\end{aligned} \tag{17}$$

where the variables are defined as below:

- $C_1$: Capacity of Capacitor1. $C_2$: Capacity of Capacitor2. $L$: Inductance of the Inductor.

- $U_1$: Voltage through Capacitor1. $U_2$: Voltage through Capacitor2.

- $I_L$: Current through Inductor. $i$: Current through the nonlinear resistor.

- $E$: the threshold voltage for the nonlinear resistor. $a$: the slope of low voltage for the nonlinear resistor. $b$: the slope of high voltage for the nonlinear resistor

- threshold: a parameter of the Chua's circuit, which equals to $E$ in this experiment.

Let

$$
\begin{aligned}
\alpha &= \frac{1}{RC_1} \\
\beta &= \frac{1}{L} \\
f(y)|_{y=x\frac{threshold}{E}} &= R\frac{threshold}{E}i(x),
\end{aligned}
\tag{18}
$$

we have the equation in the following form:

$$
\begin{aligned}
\frac{dy_1}{dt} &= \alpha[y_2 - y_1 - f(y_1)] \\
RC_2\frac{dy_2}{dt} &= y_1 - y_2 + Ry_3 \\
\frac{dy_3}{dt} &= -\beta y_1,
\end{aligned}
\tag{19}
$$

where $f(x)$ is given by

$$
f(x) = \begin{cases} k_1 x + b_1, & x < -\text{threshold} & (20a) \\ k_0 x, & -\text{threshold} < x < \text{threshold} & (20b) \\ k_1 x + b_2. & x > \text{threshold} & (20c) \end{cases}
$$

and

$$
\begin{aligned}
k_0 &= Ra \\
k_1 &= Rb \\
b_1 &= R(a-b)\text{threshold} \\
b_2 &= R(a-b)\text{threshold}.
\end{aligned}
\tag{21}
$$

The behavior of the system will be changed between chaos and non-chaos depending on the value of $R$. Each mode in eq. (20a), eq. (20b) and eq. (20c), corresponds to subsystem 1, subsystem 2 and subsystem 3 respectively. We focus on the discovery of the first equation in eq. (19) and only collect the value of $y_1$ and $y_2$. The output data from the Chua's circuit can be seen in Fig. S19.

From the true parameters in Table S25, we can compute the true coefficients of $f(x)$ to deter-

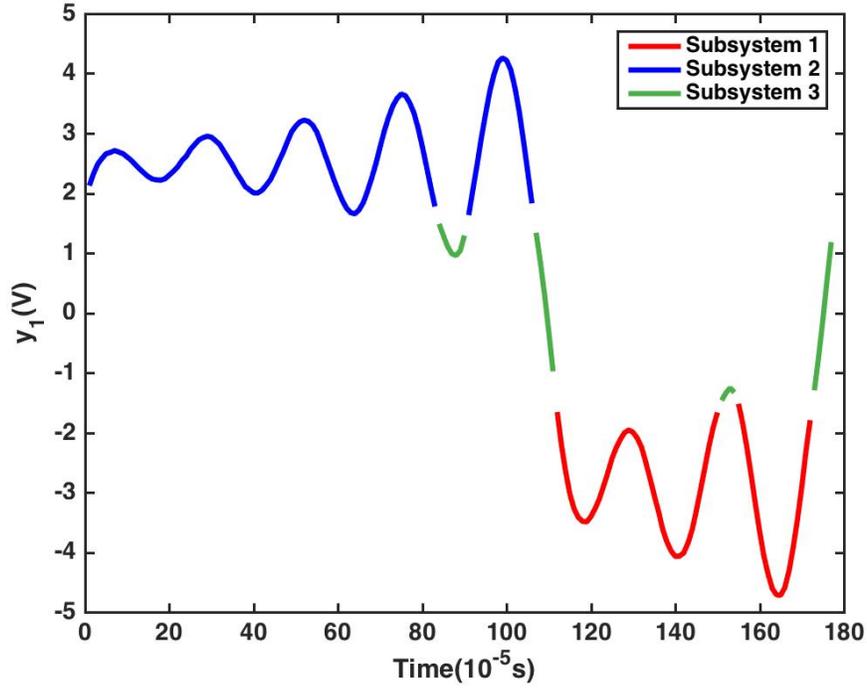

**Fig. S19.** The output trajectory of the Chua's circuit. Different colors in the trajectory specify the output generated by different subsystems.

mine $\frac{dy_1}{dt}$:

$$10^{-5}\frac{dy_1}{dt} = \begin{cases} 1.0858y_2 - 0.2115y_1 - 0.5349, & y_1 < -1.5 & (22a) \\ 1.0858y_2 + 0.1451y_1, & -1.5 < y_1 < 1.5 & (22b) \\ 1.0858y_2 - 0.1994y_1 + 0.5168. & y_1 > 1.5. & (22c) \end{cases}$$

**Table S25.** True parameters of the built Chua's circuit.

| Item | Value | Item | Value |
|---|---|---|---|
| $a$ | $-1.2309e-3$ | $c_1$ | $0.01\mu F$ |
| $b$ | $-8.743e-4$ | $c_2$ | $0.1\mu F$ |
| $b'$ | $-8.864e-4$ | $dt$ | $10^{-5}s$ |
| threshold | 1.5 | $L$ | $6.8mH$ |
| $R$ | 921 | | |

The algorithm accurately infers the form of eq. (22c) from the data as shown in Table S26.

62

**Table S26.** The identified subsystems and transition logic of Chua's circuit which contains all subsystems.

| Library $\Phi$ | $1\ y_1\ y_2$ |
|---|---|
| Identified subsystems | $10^{-5}\frac{dy_1}{dt} = 1.0889y_2 - 0.2095y_1 - 0.5622 \quad y_1 < -1.5888$ <br> $10^{-5}\frac{dy_1}{dt} = 1.0813y_2 + 0.1593y_1 \quad -1.5627 < y_1 < 1.3137$ <br> $10^{-5}\frac{dy_1}{dt} = 1.0870y_2 - 0.2128y_1 + 0.4860 \quad y_1 > 1.4627$ |
| $\lambda_z$ | 0.01 |
| $\varepsilon_z$ | 0.05 |
| $\lambda_w$ | 0.01 |
| $\varepsilon_w$ | 0.045 |
| Library $\Psi$ | $1\ y_1$ |
| $\beta$ | 0.01 |

**Table S27.** The identified subsystems and transition logic of Chua's circuit which contains all subsystems with redundant basis functions.

| Library $\Phi$ | $1\ y_1\ y_2\ e^{y_1}\ \frac{y_1}{y_2}\ \frac{\cos(0.1y_1)^2}{1+y_2^2}\ \cos(y_1+y_2)^2$ |
|---|---|
| Identified subsystems | $10^{-5}\frac{dy_1}{dt} = 1.0758y_2 - 0.2028y_1 - 0.5405 \quad y_1 < -1.4348$ <br> $10^{-5}\frac{dy_1}{dt} = 1.0793y_2 + 0.1576y_1 \quad -1.5627 < y_1 < 1.3137$ <br> $10^{-5}\frac{dy_1}{dt} = 1.0869y_2 - 0.2127y_1 + 0.4859 \quad y_1 > 1.4627$ |
| $\lambda_z$ | 0.05 |
| $\varepsilon_z$ | 0.012 |
| $\lambda_w$ | 0.01 |
| $\varepsilon_w$ | 0.044 |
| Library $\Psi$ | $1\ \ y_1\ \ y_2\ \ \sin(y_2)\cos(y_1)\ \ \frac{\frac{dy_1}{dt}}{\sin(y_1)+\frac{dy_1}{dt}}\ \ \frac{\frac{dy_1}{dt}}{y_2}\ \ \frac{dy_1}{dt}$ |
| $\beta$ | 0.01 |

## S3.10 Example 10: Autonomous Car

This example presents the results of IHYDE applying to an autonomous car built in our lab. The autonomous car consists of a body, a MK60t board, a servo motor, tow driving motors and a camera. During execution, the embedded camera captures the upcoming road layouts to check whether there is an upcoming straightaway or curve. Naturally, the car will drive faster on straightaways and slower on the curves.

Based on this design principle, we would like to design a hybrid dynamical system with two subsystems and simple transition logic to realize this goal as shown in the right panel of Figure S21. The car measures current speed by encoder and calculates the $\Delta u$, a control input to the motor. The speed control strategy is based on an incremental PI control algorithm, which is widely used in control systems. The incremental PI algorithm is developed from position PI algorithm. The position PI model is described as below and can be seen in Fig. S20. $r(t)$ represents the input of the whole system (the expected speed $v_{\text{expect}}(t)$) and $c(t)$ represents the output of the whole system (the real speed observed $v(t)$). In the figure, $u(t)$ is the output of the controller and it can

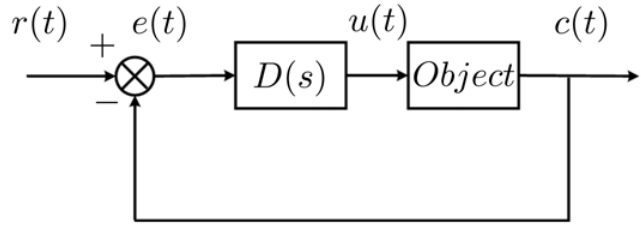

**Fig. S20.** The position PI controller structure for the autonomous car.

be calculated from $e(t)$ (where $T_I$ is the time constant for the integral control):

$$u(t) = P\left[e(t) + \frac{1}{T_I}\int_0^t e(t)dt\right]. \qquad (23)$$

In the Laplace domain, eq. (23) is equivalent to $U(s) = D(s)E(s)$, where $U(s)$ and $E(s)$ are the Laplace transform of $u(t)$ and $e(t)$ respectively, $D(s)$ represents the transfer function of the controller:

$$D(s) = \frac{U(s)}{E(s)} = P\left(1 + \frac{1}{T_I s}\right). \qquad (24)$$

Since the controller is implemented by a computer, it must be first converted to discrete time. The

integral can be approximated by

$$\int_0^t e(t)dt \approx \sum_{i=0}^{k} Te(i) \Rightarrow \frac{de(t)}{dt} \approx \frac{e(k)-e(k-1)}{T}. \tag{25}$$

So we obtain the following control law

$$u(k) = P\left[e(k) + \frac{T}{T_\text{I}}\sum_{i=0}^{k} e(i)\right]. \tag{26}$$

The position PI algorithm is usually approximated by an incremental PI algorithm:

$$\Delta u(k) \triangleq u(k) - u(k-1) = P[e(k) - e(k-1)] + Ie(k), \tag{27}$$

where $I \triangleq \frac{PT}{T_I}$. In the autonomous car example, we have

$$r(k) = v_\text{expect}(k),\ c(k) = v(k),\ e(k) = v_\text{expect}(k) - v(k). \tag{28}$$

Here $v_\text{expect}(k)$ is the expected velocity depending on whether there is an upcoming straightaway or curve from the camera. We set up a faster velocity on straightaways and slower one on the curves. Substituting $e(k)$ into the equation, we obtain

$$\Delta u(k) = P[v(k-1) - v(k)] + P[v_\text{expect}(k) - v_\text{expect}(k-1)] + I(v_\text{expect}(k) - v(k)). \tag{29}$$

When the car changes its expected velocity, this could lead to a more complicated hybrid dynamical system than we would design as shown in Figure S21. This side effect is due to the abrupt switching and discretization. In practice, we normally neglect these subsystems in the modeling, analysis and design. The flow chart of the PI control algorithm is shown in Fig. S22.

Next, we demonstrate how IHYDE can help in the design process. In the first experiment, the autonomous car failed to drive through the track. We collected the experimental data and used IHYDE to discover the failed system. We compared the discovered system model with the to-be designed one and found an implementation error that led the system to failure. We expected a higher speed when the car is running in a straight line and a lower speed while it is running on a curve. The model from the failed experiments showed that the transition logistics should be reversed as shown in Fig. S20 and Table S28. We fixed the bug and as a result the autonomous car was able to run through the track. Finally, as a validation, we collected the data and repeated the modeling process in Table S29 and Table S30.

In summary, IHYDE successfully reverse engineered the control strategy of the CPS. Addi-

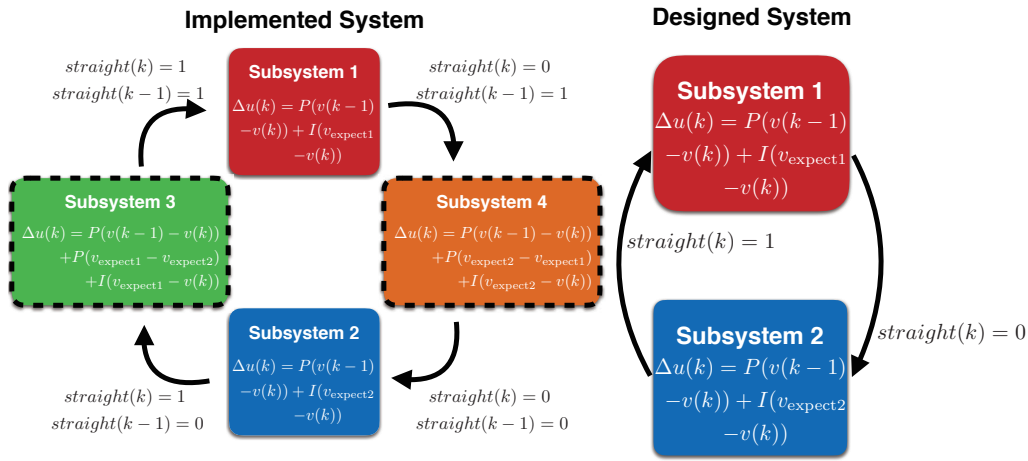

**Fig. S21.** Left: a more complicated hybrid dynamical system model due to discretization and switching. Right: the correct hybrid dynamical system model that we would like to design.

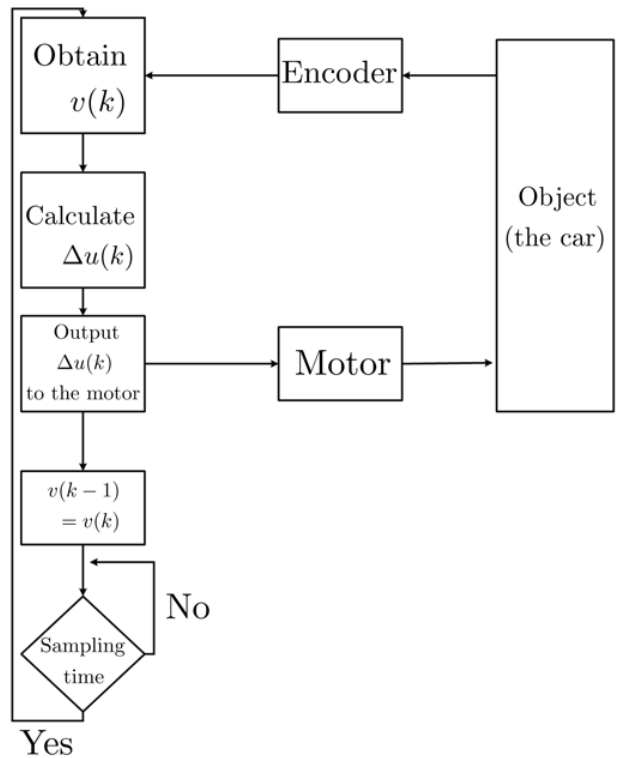

**Fig. S22.** The flow chart of the PI algorithm.

tionally, we deliberately swapped the straightway and curve speeds to mimic a software bug. The modeled system immediately pinpointed the location of the faulty software and yielded important information for debugging the system.

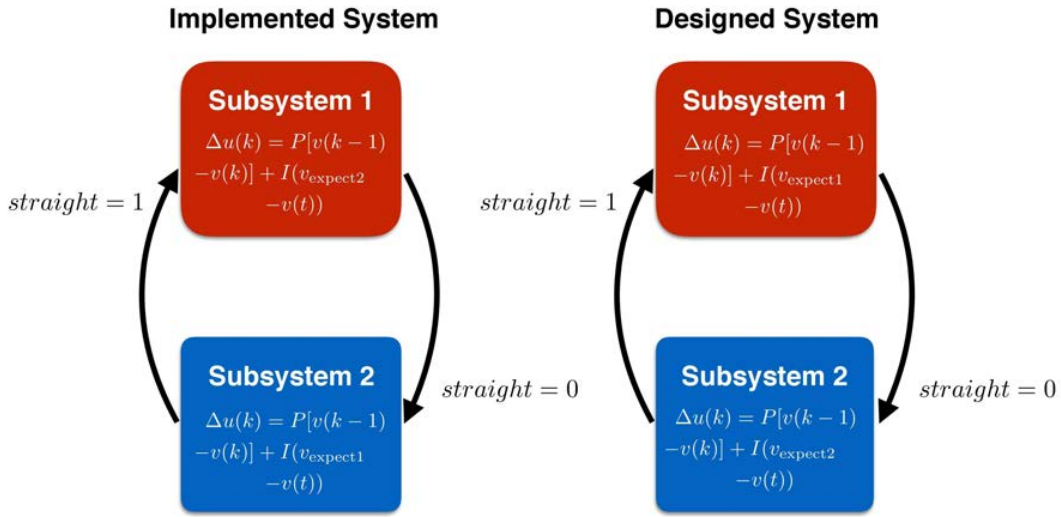

**Fig. S23.** The IHYDE pinpoints the implementation error that leads to a failure in the autonomous car experiment. Left: the identified model from experimental data. Right: the designed system. The two subsystems are swapped around due to a design bug.

**Table S28.** The identified transition logic of autonomous car testbed.

| System | Straightaway | Curve |
|---|---|---|
| Straightaway | | straight $< 0.3318$ |
| Curve | straight $> 0.6072$ | |
| Library $\Psi$ | 1 straight $\sin(v(k))$ $\cos(v(k))$ $\tan(v(k))$ $\frac{v(k-1)-v(k-4)}{v(k-2)}$ $v(k-1)\tan(v(k-3))$ | |
| $\beta$ | 0.001 | |

**Table S29.** The identified result and details of autonomous car testbed.

| Library $\Phi$ | \multicolumn{2}{c}{1 $v(k)$ $v(k-1)$} | |
|---|---|---|
| Speed control strategy | Straightaway | Curve |
| True strategy | $\Delta u(k) = 9.5(380 - v(k))$ $+48(v(k-1) - v(k))$ | $\Delta u(k) = 9.5(280 - v(k))$ $+48(v(k-1) - v(k))$ |
| True times | 147 | 249 |
| Identified strategy | $\Delta u(k) = 9.4957(379.9550 - v(k))$ $+47.9742(v(k-1) - v(k))$ | $\Delta u(k) = 9.4961(279.9469 - v(k))$ $+47.9927(v(k-1) - v(k))$ |
| Identified switching times | 147 | 249 |
| $\lambda_z$ | 0.1 | |
| $\varepsilon_z$ | 100 | |
| $\lambda_w$ | $1e-5$ | |
| $\varepsilon_w$ | 8 | |

**Table S30.** The identified result and details of autonomous car testbed with redundant basis functions.

| Library $\Phi$ | all the polynomial combinations of $v(k),\ldots,v(k-4)$ to fourth order | |
|---|---|---|
| Speed control strategy | Straightaway | Curve |
| True strategy | $\Delta u(k) = 9.5(380 - v(k))$ <br> $+48(v(k-1) - v(k))$ | $\Delta u(k) = 9.5(280 - v(k))$ <br> $+48(v(k-1) - v(k))$ |
| True times | 147 | 249 |
| Identified strategy | $\Delta u(k) = 9.4957(379.9554 - v(k))$ <br> $+47.9741v(k-1) - v(k))$ | $\Delta u(k) = 9.4959(279.9465 - v(k))$ <br> $+47.9888(v(k-1) - v(k))$ |
| Identified switching times | 147 | 249 |
| $\lambda_z$ | 0.01 | |
| $\varepsilon_z$ | 100 | |
| $\lambda_w$ | $1e-5$ | |
| $\varepsilon_w$ | 8 | |

## S3.11 Example 11: Non-hybrid Dynamical Systems

We also tested the proposed method to non-hybrid dynamical systems using datasets in *[13]* to illustrate the applicability of IHYDE. The results are summarized in Table S31 and Table S32. These tables show that the proposed subsystem identification algorithm unifies previous results for the discovery of nonhybrid dynamical systems, such as examples in *[13]*.

**Table S31. The identified results using datasets in *[13]*.** Five prototypical systems were examined. We used the actual systems to simulate datasets and then used IHYDE to reverse engineering these systems from data.

| Examples | Actual systems | Discovered systems |
|---|---|---|
| Linear 2D | $\frac{d}{dt}\begin{bmatrix}x\\y\end{bmatrix} = \begin{bmatrix}-0.1 & 2\\-2 & -0.1\end{bmatrix}\begin{bmatrix}x\\y\end{bmatrix}$ | $\frac{d}{dt}\begin{bmatrix}x\\y\end{bmatrix} = \begin{bmatrix}-0.0993 & 2.0054\\-2.0004 & -0.1048\end{bmatrix}\begin{bmatrix}x\\y\end{bmatrix}$ |
| Cubic 2D | $\frac{d}{dt}\begin{bmatrix}x\\y\end{bmatrix} = \begin{bmatrix}-0.1 & 2\\-2 & -0.1\end{bmatrix}\begin{bmatrix}x^3\\y^3\end{bmatrix}$ | $\frac{d}{dt}\begin{bmatrix}x\\y\end{bmatrix} = \begin{bmatrix}-0.1015 & 2.0005\\-2.0010 & -0.1002\end{bmatrix}\begin{bmatrix}x^3\\y^3\end{bmatrix}$ |
| Linear 3D | $\frac{d}{dt}\begin{bmatrix}x\\y\\z\end{bmatrix} = \begin{bmatrix}-0.1 & 2 & 0\\-2 & -0.1 & 0\\0 & 0 & -0.3\end{bmatrix}\begin{bmatrix}x\\y\\z\end{bmatrix}$ | $\frac{d}{dt}\begin{bmatrix}x\\y\\z\end{bmatrix} = \begin{bmatrix}-0.0992 & 2.0002 & 0\\-1.9999 & -0.0991 & 0\\0 & 0 & -0.2983\end{bmatrix}\begin{bmatrix}x\\y\\z\end{bmatrix}$ |
| Lorenz system | $\dot{x} = 10y - 10x$ <br> $\dot{y} = 28x - xz - y$ <br> $\dot{z} = xy - 2.6667z$ | $\dot{x} = 10.0060y - 9.9968x$ <br> $\dot{y} = 27.9480x - 0.9957xz - 0.9954y$ <br> $\dot{z} = 1.0010xy - 2.6673z$ |
| Logistic map | $x_{k+1} = \mu_k x_k(1 - x_k)$ <br> $\mu_{k+1} = \mu_k$ | $x_{k+1} = \mu_k x_k(1.0005 - 1.0006 x_k)$ <br> $\mu_{k+1} = 1.0000\mu_k$ |

**Table S32. The identified results using datasets in *[13]*.** The tuning parameters are presented for these prototypical examples.

| Examples | Noise | $\lambda_z$ | $\varepsilon_z$ | $\lambda_w$ | $\varepsilon_w$ | Numbers of points |
|---|---|---|---|---|---|---|
| Linear 2D | 0.05 | 1 | $1e-4$ | $2e-3$ | 0.2 | 1001 |
| Cubic 2D | 0.05 | 1 | $1e-4$ | $2e-3$ | 0.2 | 1001 |
| Linear 3D | 0.01 | 1 | $1e-4$ | $2e-3$ | 0.2 | 1001 |
| Lorenz system | 1 | 1 | $1e-4$ | $1e-3$ | 4 | 1000 |
| Logistic map | 0.01 | 1 | $1e-4$ | $2e-3$ | 0.2 | 990 |

# S4 Discussion

The IHYDE algorithm has been tested in a number of examples. As the number of basis functions and the amount of noise increases, the algorithm is eventually unable to identify the actual model. Although it can fit data very well, it usually obtains more complex models than the true ones. This is actually a typical problem in system identification *[28]*. When the data is not informative, it leads to non-identifiability issues, i.e., there will exist multiple hybrid dynamical systems that can produce the same data, which prevents the proposed IHYDE algorithm from finding the true system.

## S4.1 Example 1

Consider the following linear system with unknown parameters $k_1$ and $k_2$

$$\frac{d}{dt}\begin{bmatrix} x_1 \\ x_2 \end{bmatrix} = \begin{bmatrix} k_1 & 1+k_2 \\ 0 & k_1+k_2 \end{bmatrix} \begin{bmatrix} x_1 \\ x_2 \end{bmatrix} + \begin{bmatrix} 0 \\ 1 \end{bmatrix} u$$

$$y = \begin{bmatrix} 0 & 1 \end{bmatrix} \begin{bmatrix} x_1 \\ x_2 \end{bmatrix}.$$

(30)

The observed output is plotted as follows in Fig. S24 (the system is stimulated by an impulse input, i.e., $u(t) = \delta(t)$ where $\delta(\cdot)$ is the Dirac delta function):

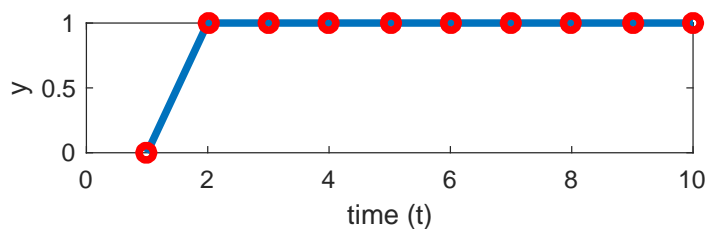

**Fig. S24.** The observed output of eq. (30) when a Dirac delta function is applied to stimulate the system.

However, any $k_1, k_2$ with $k_1 + k_2 = 0.8$ produces the same input-output data. For example, the actual system

$$\frac{d}{dt}\begin{bmatrix} x_1 \\ x_2 \end{bmatrix} = \begin{bmatrix} 0.4 & 1.4 \\ 0 & 0.8 \end{bmatrix} \begin{bmatrix} x_1 \\ x_2 \end{bmatrix} + \begin{bmatrix} 0 \\ 1 \end{bmatrix} u$$
$$y = \begin{bmatrix} 0 & 1 \end{bmatrix} \begin{bmatrix} x_1 \\ x_2 \end{bmatrix},$$
(31)

and

$$\frac{d}{dt}\begin{bmatrix} x_1 \\ x_2 \end{bmatrix} = \begin{bmatrix} 0.3 & 1.5 \\ 0 & 0.8 \end{bmatrix} \begin{bmatrix} x_1 \\ x_2 \end{bmatrix} + \begin{bmatrix} 0 \\ 1 \end{bmatrix} u$$
$$y = \begin{bmatrix} 0 & 1 \end{bmatrix} \begin{bmatrix} x_1 \\ x_2 \end{bmatrix}$$
(32)

are indistinguishable from the input-output data alone. Hence, without more information, the true parameters cannot be identified using any methods.

## S4.2 Example 2

The previous example demonstrates that, when the parameterization is not identifiable, no algorithm is able to identify the correct parameters. Next, we shall demonstrate another example where, even though the subsystem is identifiable, the data is not informative enough. For example, some of the logic transitions never occur. Consider the following hybrid dynamical system in Fig. S25. If the system starts at initial condition $y(0) = 18$, then it always stays in subsystem 1. Hence, with the data generated, no algorithm is able to identify the complete hybrid dynamical system.

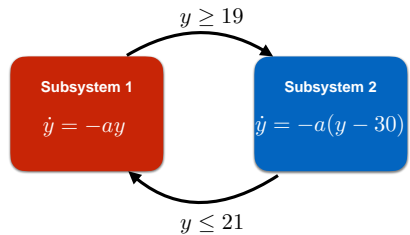

**Fig. S25.** A counterexample of a constructed hybrid dynamical system that is not able to be identified from data.

# A  Extension of the Proposed IHYDE Algorithm

Both subsystems identification algorithms require solving a number of $\ell_1$ regularized optimization problems. In practice, we introduced an alternative method to formulate the optimization problem by introducing a sparse prior and re-deriving the process using Bayesian statistics. This can lead to recursive $\ell_1$ regularized optimizations (shown below). This section is a brief introduction to the results in *[10]*, for the completeness of the paper, we present it here.

In the following, we shall first interpret the $\ell_1$-minimization problem from a Bayesian perspective and show that the regularized least square problem is equivalent to introducing a sparsity-inducing prior in the Bayesian framework. We demonstrate that the corresponding cost function is concave both in the hyperparameter space and in the weight space. Finally, we analyze the cost function and propose algorithms to relax the concave optimization problem into an iterative convex optimization problem based on the reweighted $\ell_1$-minimization algorithm.

Bayesian modeling treats all unknowns as stochastic variables with certain probability distributions. Consider a generalized model $y = \Theta w + \xi$ with $\xi$ being a vector containing stochastic variables, it's assumed that the stochastic variables in $\xi$ are i.i.d. Gaussian distributed with $\xi \sim \mathcal{N}(0, \sigma^2 I)$. Then the likelihood of the output $y$ given a weight $w$ is

$$P(y|w) = \mathcal{N}(y|\Theta w, \sigma^2 I) \propto \exp\left[-\frac{1}{2\sigma^2}\|y - \Theta w\|_2^2\right]. \tag{33}$$

Note that obtaining the maximum likelihood estimates for $w$ in eq. (33) is equivalent to searching for a minimal $\ell_2$-norm solution $w$, i.e., $\min \|y - \Theta w\|_2^2$. Given the likelihood function in eq. (33) and specifying a prior of $w$

$$P(w) = \prod_j P(w_j), \tag{34}$$

we compute the *posterior distribution* over $w$ via Bayes' rule:

$$P(w|y) = \frac{likelihood \times prior}{normalizing\ factor} = \frac{P(y|w)P(w)}{P(y)} = \frac{P(y|w)P(w)}{\int P(y|w)P(w)dw}. \tag{35}$$

Let $P(w)$ be defined as

$$P(w) \propto \exp\left[-\frac{1}{2}g(w)\right] = \exp\left[-\frac{1}{2}\sum_j g(w_j)\right], \tag{36}$$

where $g(w_j)$ is an arbitrary penalty function of $w_j$. Combining $P(y|w)$ and $P(w)$, we get the

72

posterior distribution

$$P(w|y) \propto P(y|w)P(w). \tag{37}$$

Based on the fact that the posterior distributions $P(w|y)$ and $P(y)$ are independent of $w$, we can derive the following cost function

$$\mathscr{L}(w) \triangleq -2\log P(w|y) \propto \frac{1}{\sigma^2}\|y - \Theta w\|_2^2 + g(w). \tag{38}$$

We then formulate a maximum a posteriori (MAP) estimate $w$ to $y = \Theta w + \xi$:

$$w_{MAP} = \arg\max_w P(w|y) = \arg\min_w \{\|y - \Theta w\|_2^2 + \sigma^2 g(w)\}. \tag{39}$$

From a Bayesian viewpoint, MAP estimation is equivalent to a regularized least square problem. Setting $g(w) = \|w\|_1$ and $\lambda_z = \sigma^2$, we recover the $\ell_1$ regularized optimization in eq. (6). We may also consider *super-Gaussian* priors, which yield a lower bound for the priors $P(w_j)$ *[41]*. More specifically, if we define $\gamma \triangleq [\gamma_1, \ldots, \gamma_N]^T \in \mathbb{R}_+^N$, we can represent the prior in eq. (34) in the following relaxed (variational) form *[40]*:

$$P(w) = \prod_{j=1}^n P(w_j), \ P(w_j) = \max_{\gamma_j > 0} \mathscr{N}(w_j|0, \gamma_j)\varphi(\gamma_j), \tag{40}$$

where $\varphi(\gamma_j)$ is a nonnegative function which is treated as a hyperprior with $\gamma_j$ being associated hyperparameters. Throughout, we call $\varphi(\gamma_j)$ the potential function. This Gaussian relaxation is possible if and only if $\log P(\sqrt{w_j})$ is concave on $(0, \infty)$ shown in *[40]*.

For a fixed $\gamma = [\gamma_1, \ldots, \gamma_N]^T$, define a relaxed prior which is a joint probability distribution of $w$ and $\gamma$

$$P(w; \gamma) = \prod_j \mathscr{N}(w_j|0, \gamma_j)\varphi(\gamma_j) = P(w|\gamma)P(\gamma), \tag{41}$$

where $P(w|\gamma) \triangleq \prod_j \mathscr{N}(w_j|0, \gamma_j), P(\gamma) \triangleq \prod_j \varphi(\gamma_j)$. Since is $P(y|w)$ is Gaussian in eq. (33), we can get a relaxed Gaussian posterior

$$P(w|y, \gamma) = \frac{P(y|w)P(w; \gamma)}{\int P(y|w)P(w; \gamma)dw} = \mathscr{N}(m_w, \Sigma_w), \tag{42}$$

where

$$m_w = \Gamma \Theta^T (\sigma^2 I + \Theta \Gamma^{-1} \Theta^T)^{-1} y, \tag{43}$$

$$\Sigma_w = \Gamma - \Gamma \Theta^T (\sigma^2 I + \Theta \Gamma^{-1} \Theta^T)^{-1} \Theta, \tag{44}$$

with $\Gamma \triangleq \text{diag}[\gamma]$. Next we shall choose the most appropriate $\gamma = \hat{\gamma} = [\hat{\gamma}_1, \ldots, \hat{\gamma}_N]^T$ to maximize $\prod_j \mathcal{N}(w_j|0, \gamma_j)\varphi(\gamma_j)$ such that $P(w|y, \hat{\gamma})$ can be a good relaxation to $P(w|y)$. Using the product rule for probabilities, we can write the full posterior

$$P(w, \gamma|y) \propto P(w|y, \gamma)P(\gamma|y) = \mathcal{N}(m_w, \Sigma_w) \times \frac{P(y|\gamma)P(\gamma)}{P(y)}. \tag{45}$$

Since $P(y)$ is independent of $\gamma$, the quantity

$$P(y|\gamma)P(\gamma) = \int P(y|w)P(w|\gamma)P(\gamma)dw$$

is the prime target for variational methods *[42]*. This quantity is known as evidence or marginal likelihood. One way to select $\hat{\gamma}$ is to choose it as the minimizer of the sum of the misaligned probability mass, e.g.,

$$\begin{aligned}
\hat{\gamma} &= \arg\min_{\gamma \geq 0} \int P(y|w)|P(w) - P(w;\gamma)|dw \\
&= \arg\max_{\gamma \geq 0} \int P(y|w) \prod_{j=1}^n \mathcal{N}(w_j|0, \gamma_j)\varphi(\gamma_j)dw.
\end{aligned} \tag{46}$$

Since $P(w)$ is not a function of $\gamma$, and from the variational representation in eq. (40), we can remove the absolute value. The procedure in eq. (46) is referred to as evidence maximization or type-II maximum likelihood *[43]*. It means that the marginal likelihood can be maximized by selecting the most probable hyperparameters able to explain the observed data. Once $\hat{\gamma}$ is computed, an estimate of the unknown weights can be obtained by setting $\hat{w}$ to the posterior mean in eq. (43)

$$\hat{w} = \mathbb{E}(w|y; \hat{\gamma}) = \hat{\Gamma}\Theta^T(\sigma^2 I + \Theta\hat{\Gamma}\Theta^T)^{-1}y. \tag{47}$$

We shall now propose an algorithm to compute $\hat{\gamma}$ in eq. (46). Then, from this computed $\hat{\gamma}$ we can obtain an estimation of the posterior mean $\hat{w}$ in eq. (47) based on the reweighted $\ell_1$-minimization algorithm.

**Theorem 1.** *[10] The optimal hyperparameters $\hat{\gamma}$ in (46) can be achieved by minimizing the following cost function*

$$\mathscr{L}_\gamma(\gamma) = \log|\sigma^2 I + \Theta\Gamma^{-1}\Theta^T| + y^T(\sigma^2 I + \Theta\Gamma^{-1}\Theta^T)^{-1}y + \sum_{j=1}^N p(\gamma_j), \tag{48}$$

*where $p(\gamma_j) = -2\log\varphi(\gamma_j)$.*

Then, from eq. (47), we compute the posterior mean to get an estimate to $w$:

$$\hat{w} = \mathbb{E}(w|y;\hat{\gamma}) = \hat{\Gamma}\Theta^T(\sigma^2 I + \Theta\hat{\Gamma}\Theta^T)^{-1}y$$

where $\hat{\Gamma} = \text{diag}[\hat{\gamma}]$.

The data-dependent term can be re-expressed as

$$\begin{aligned}
y^T\left(\sigma^2 I + \Theta\Gamma^{-1}\Theta^T\right)^{-1}y &= \sigma^{-2}y^T y - \sigma^{-2}y^T\Theta\Sigma_w\Theta^T\sigma^{-2}y \\
&= \sigma^{-2}y^T(y - \Theta m_w) \\
&= \sigma^{-2}\|y - \Theta m_w\|_2^2 + m_w^T\Sigma_w^{-1}m_w - \sigma^{-2}m_w^T\Theta^T\Theta m_w \\
&= \sigma^{-2}\|y - \Theta m_w\|_2^2 + m_w^T\Gamma m_w \\
&= \min_w\{\frac{1}{\sigma^2}\|y - \Theta w\|_2^2 + w^T\Gamma w\}.
\end{aligned} \quad (49)$$

From eq. (49), we can create a strict upper bounding auxiliary function $\mathscr{L}_{\gamma,w}(\gamma,w)$ on $\mathscr{L}_\gamma(\gamma)$ in eq. (46),

$$\begin{aligned}
\mathscr{L}_{\gamma,w}(\gamma,w) &\triangleq \langle\gamma^*,\gamma\rangle - h^*(\gamma^*) + y^T\left(\sigma^2 I + \Theta\Gamma^{-1}\Theta^T\right)^{-1}y \\
&= \frac{1}{\sigma^2}\|y - \Theta w\|_2^2 + \sum_j\left(\frac{w_j^2}{\gamma_j} + \gamma_j^*\gamma_j\right) - h^*(\gamma^*).
\end{aligned} \quad (50)$$

Then we define the terms excluding $h^*(\gamma^*)$ in eq. (50) as

$$\mathscr{L}_{\gamma^*}(\gamma,w) \triangleq \frac{1}{\sigma^2}\|y - \Theta w\|_2^2 + \sum_j\left(\frac{w_j^2}{\gamma_j} + \gamma_j^*\gamma_j\right). \quad (51)$$

For a fixed $\gamma^*$, we notice that $\mathscr{L}_{\gamma^*}(\gamma,w)$ is jointly convex in $w$ and $\gamma$ and can be globally minimized by solving over $\gamma$ and then $w$. Since $w_j^2/\gamma_j + \gamma_j^*\gamma_j \geq 2w_j\sqrt{\gamma_j^*}$, for any $w$, $\gamma_j = |w_j|/\sqrt{\gamma_j^*}$ minimizes $\mathscr{L}_{\gamma^*}(\gamma,w)$.

The next step is to find a $\hat{w}$ that minimizes $\mathscr{L}_{\gamma^*}(\gamma,w)$. When $\gamma_j = |w_j|/\sqrt{\gamma_j^*}$ is substituted into $\mathscr{L}_{\gamma^*}(\gamma,w)$, $\hat{w}$ can be obtained by solving the following weighted convex $\ell_1$-minimization problem

$$\begin{aligned}
\hat{w} &= \arg\min_w\{\|y - \Theta w\|_2^2 + 2\sigma^2\sum_{j=1}^N u_j|w_j|\} \\
&= \arg\min_w\{\|y - \Theta w\|_2^2 + 2\sigma^2\sum_{j=1}^N\sqrt{\gamma_j^*}|w_j|\}.
\end{aligned} \quad (52)$$

where $\sqrt{\gamma_j^*}$ are the weights. We can then set

$$\gamma_j = \frac{|\hat{w}_j|}{\sqrt{\gamma_j^*}}, \forall j, \tag{53}$$

and, as a consequence, $\mathscr{L}_{\gamma^*}(\gamma, w)$ will be minimized for any fixed $\gamma^*$.

Consider $\mathscr{L}_{\gamma,w}(\gamma, w)$ in eq. (50) again, for any fixed $\gamma$ and $w$, the tightest bound can be obtained by minimizing over $\gamma^*$. The tightest value of $\gamma^* = \hat{\gamma}^*$ equals the slope at the current $\gamma$ of the function $h(\gamma) \triangleq \log|\sigma^2 I + \Theta \Gamma^{-1} \Theta^T| + \sum_{j=1}^{N} p(\gamma_j)$. Using basic principles of convex analysis, we then obtain the following analytic form for the optimizer $\gamma^*$:

$$\begin{aligned}
\hat{\gamma}^* &= \nabla_\gamma \left( \log|\sigma^2 I + \Theta \Gamma^{-1} \Theta^T| + \sum_{j=1}^{N} p(\gamma_j) \right) \\
&= \text{diag}\left[ \Theta^T \left( \sigma^2 I + \Theta \Gamma^{-1} \Theta^T \right)^{-1} \Theta \right] + p'(\gamma),
\end{aligned} \tag{54}$$

where $p'(\gamma) = [p'(\gamma_1), \ldots, p'(\gamma_N)]^T$.

The algorithm is then based on successive iterations of eq. (52), eq. (53) and eq. (54) until it converges to $\hat{\gamma}$. We can then compute the posterior mean and covariance as in eq. (43) and eq. (44)

$$\begin{aligned}
\hat{w} &= \hat{\Gamma} \Theta^T (\sigma^2 I + \Theta \hat{\Gamma} \Theta^T)^{-1} y \\
\Sigma_{\hat{w}} &= \hat{\Gamma} - \hat{\Gamma} \Theta^T (\sigma^2 I + \Theta \hat{\Gamma} \Theta^T)^{-1} \Theta
\end{aligned} \tag{55}$$

where $\hat{\Gamma} = \text{diag}[\hat{\gamma}]$. The above described procedure is summarized in Algorithm 3.

**Algorithm 3** Reweighted $\ell_1$-minimization

**Data:** $y$ and $\Theta$

**Result:** Posterior mean for $w$.

**Step 1.** *Set iteration count $k$ to zero and initialize each $u_j^{(0)} = \sqrt{\gamma_j^*}$, with randomly chosen initial values for $\gamma_j^*$, $\forall j$, e.g., with $\gamma_j^* = 1$, $\forall j$. Set $k = 1$.*

**Step 2.** *At the $k^{th}$ iteration, solve the reweighted $\ell_1$-minimization problem*

$$\hat{w}^{(k)} = \arg\min_w \{\|y - \Theta w\|_2^2 + 2\sigma^2 \sum_{j=1}^N u_j^{(k)} |w_j|\}.$$

**Step 3.** *Compute $\gamma_j^{(k)}$*

$$\gamma_j^{(k)} = \frac{|\hat{w}_j^{(k)}|}{\sqrt{\gamma_j^{*(k)}}}, \forall j,$$

**Step 4.** *Update $\hat{\gamma}^{*(k+1)}$ using eq. (54)*

$$\hat{\gamma}^{*(k+1)} = \text{diag}\left[\Theta^T \left(\sigma^2 I + \Theta \Gamma^{(k)} \Theta^T\right)^{-1} \Theta\right] + p'(\gamma^{(k)}).$$

**Step 5.** *Update weights $u_j^{(k+1)}$ for the $\ell_1$-minimization at the next iteration $u_j^{(k+1)} = \sqrt{\hat{\gamma}_j^{*(k+1)}}$.*

**Step 6.** *Set $k = k + 1$. Iterate Steps 2 to 5 until convergence to some $\hat{\gamma}$.*

**Step 7.** *Compute $\hat{w} = \mathbb{E}(w|y;\hat{\gamma}) = \hat{\Gamma}\Theta^T(\sigma^2 I + \Theta\hat{\Gamma}\Theta^T)^{-1}y$.*

# B  User's Manual of the Code

Identification of hybrid dynamical systems (IHYDE) is a open-source Matlab toolbox for automating the mechanistic modeling of hybrid dynamical systems from observed data. IHYDE has much lower computational complexity comparing with genetic algorithms *[35]*, enabling its application to real-world CPS problems. IHYDE implements the clustering-based algorithms described in the Data-driven Discovery of Cyber Physical Systems. It can also be used, positively, for the creation of guidelines for designing new CPSs. IHYDE uses routines of the CVX and SLR toolboxes for constructing and solving disciplined convex programs (DCPs).

Download the latest version of IHYDE toolbox in a directory and add its path (and the path of the subdirectories) to the Matlab path. The IHYDE toolbox consists of directories listed in Table S33.

**Table S33.** Directories in the IHYDE toolbox.

| Directories | Description |
|---|---|
| /IHYDE | main functions and examples |
| /IHYDE/docs | pdf and html documentation |
| /IHYDE/data | datasets we used in the paper |
| /IHYDE/tools | functions for IHYDE |

Table S34, Table S35, Table S36 and Table S37 contain the introduction of IHYDE's API.

**Table S34.** The introduciton of function library which can construct library for further identification.

| Function library | Description |
|---|---|
| yin | an $m$ by $n$ matrix which contains time-course input-output data. In here, $m$ is the sample number, and $n$ is the number of variables. |
| polyorder | used to construct the polynomial of the highest order (up to fifth order). |
| basis_function | add more basis functions. It can be turned off, if basis_function.work set as 'off'. |
| Phi | constructed dictionary matrix $\Phi$. |

**Table S35.** The introduction of function ihyde. The ihyde can be used to identify each subsystem.

| Function ihyde | Description |
|---|---|
| parameter.y | the output data. |
| parameter.normalize_y | set to 1 if $y$ need to be normalized. |
| parameter.max_s | the max number of subsystems that could be identified by IHYDE. |
| parameter.epsilon | a 1 by 2 vector. For finding new subsystems. |
| parameter.lamdba | a 1 by 2 vector. For the identification of a new subsystems. |
| parameter.MAXITER | the max number of iterations that the sparsesolver function solves. |
| result. idx_sys | the index of each subsystem. |
| result.sys | the model of each subsystem. |
| result.theta | $z$ of each identified subsystem. |
| result.error | the fitting error. |

**Table S36.** The introduction of function finetuning. Based on the minimum error principle, it finetunes the result from ihyde and outputs the final result.

| Function finetuning | Description |
|---|---|
| result.lambda | the trading-off parameter $\lambda$ of the `sparsesolver` function. Parameter.lamdba(2) is set as the default value. |
| result.epsilon | the threshold in finetuning. Parameter.epsilon(2) is set as the default value. |
| result.threshold | the threshold for subsystem clustering. |
| final_result.idx | the index of each subsystem. |
| final_result.sys | the model of each subsystem. |
| final_result.allerror | the error which compared with the true output. |

**Table S37.** The introduction of function ihydelogic.

| Function ihydelogic | Description |
|---|---|
| para_log.Phi2 | constructed dictionary matrix $\Psi$ for inferring transition logic of each subsystem. |
| para_log.idx_sys | the index of each subsystem. |
| para_log.beta | the tradeoff parameter in the $\ell_1$ sparse logistic regression. |
| para_log.y | the output data. |
| para_log.normalize | set to 1 if Phi2 need to be normalized. |

To quickly familiarize with IHYDE, examples are in the directory /IHYDE/examples. These .m files can also be used as templates for other experiments. We shall use the autonomous car example to explain the code. First, we load the data.

```matlab
clear all,close all,clc
addpath('./tools');
addpath('./data');
basis_function.work='off';
data=load('normal_car.mat');   %%  load Data
index = 1000:1400;
flag = data.flag(index);    % 1:straghtway   0:curve
dy = data.dy(index);
v = data.v(index)/10;
```

```matlab
memory = 4; %The data covers the time from k-4 to k
polyorder = 4; % The highest order of the polynomial is 4
    order
usesine  = 0;%no sin and cos
A= library(v,polyorder,usesine,memory,basis_function);%
    make a library
A = A(memory+2:end,:);
dy = dy(memory+2:end); %dpwm_{k}
flag = flag(memory+2:end); %flag_{k}

v_k1 = v(memory+1:end-1,:);  % v_{k-1}
v_k2 = v(memory:end-2,:);% v_{k-2}
v_k3 = v(memory-1:end-3,:);% v_{k-3}
v = v(memory+2:end,:);  % v_{k}
```

Then, we initialize the parameters and identify the systems by function `ihyde`.

```matlab
parameter.MAXITER = 5;    %the iter for  the sparsesolver
    algorithm
parameter.max_s = 20;            % the max number of
    subsystems
parameter.epsilon = [100  8];   %  the fist element in
    lambda is  epsilon_z and the second is epsilon_w
parameter.Phi = A;    % the library
parameter.y = dy;  %dpwm
parameter.normalize_y = 1;  % normalize:1    unnormalize
    :0
```

80

```matlab
            [result]=ihyde(parameter);      % inferring  subsystems
```

Function `ihyde` will return a preliminary identified result which contains the details of subsystems. Since we want to get a better result based on the minimum error principle, we use function `finetuning` to fine-tune the results.

```matlab
            result.epsilon = parameter.epsilon(2);% use epsilon_w as
                the epsilon in finetuning
            result.lambda = parameter.lambda(2);% use lambda_w as the
                 epsilon in finetuning
            result.threshold = [0.05];     %set a threshold for
                clustering
            final_result  = finetuning(result);    % finetuning each
                subsystems
            sys = final_result.sys;  % get the identified subsystems
            idx_sys = final_result.idx;% get the index of each
                subsystems
```

The code for inferring transition logic between subsystems is shown below.

```matlab
            Phi2 = [ones(size(flag)) flag 1./v sin(v) cos(v) v.^2
                v_k1./v_k2 v_k3.^2 ];% library for inferring
                transition logic between subsystems.

            para_log.idx_sys = idx_sys;
            para_log.beta= 0.5;   % the tradeoff of l1-sparse
                logistic regression
            para_log.y = dy;
            para_log.Phi2 = Phi2;

            [syslogic,labelMat,data] = ihydelogic(para_log);
```

The identified results are saved in `sys`, `idx_sys` and `syslogic`.



\end{document}